\documentclass[sn-chicago]{sn-jnl}% Chicago-based Humanities Reference Style
\usepackage[utf8]{inputenc}

\usepackage{csvsimple} % added myself

\usepackage{lscape} % added myself

\usepackage{rotating} % added myself

\usepackage{adjustbox}

\usepackage{longtable} % added by Tian

\usepackage{makecell} % added myself

\usepackage{amssymb}
\usepackage{xcolor}

\usepackage{graphicx}
\usepackage{tabularx}
\usepackage{breakurl}
\usepackage{natbib}
\usepackage{multirow}

\usepackage[normalem]{ulem}
\useunder{\uline}{\ul}{}

\usepackage{scalerel}
\usepackage{dirtree}
\usepackage{multicol}
\usepackage{mathrsfs}

\usepackage{listings}
\usepackage{color}
\definecolor{gray}{rgb}{0.4,0.4,0.4}
\definecolor{darkblue}{rgb}{0.0,0.0,0.6}
\definecolor{cyan}{rgb}{0.0,0.6,0.6}

\lstdefinelanguage{XML}
{
  morestring=[b]",
  morestring=[s]{>}{<},
  morecomment=[s]{<?}{?>},
  stringstyle=\color{black},
  identifierstyle=\color{darkblue},
  keywordstyle=\color{cyan},
  morekeywords={xmlns,version,type}% list your attributes here
}

\lstdefinelanguage{json}
{
  basicstyle=\ttfamily,
  columns=fullflexible,
  showstringspaces=false,
  commentstyle=\color{gray}\upshape,
  string=[s]{"}{"},
  stringstyle=\color{darkblue},
  literate=
    *{0}{{{\color{darkblue}0}}}{1}
     {1}{{{\color{darkblue}1}}}{1}
     {2}{{{\color{darkblue}2}}}{1}
     {3}{{{\color{darkblue}3}}}{1}
     {4}{{{\color{darkblue}4}}}{1}
     {5}{{{\color{darkblue}5}}}{1}
     {6}{{{\color{darkblue}6}}}{1}
     {7}{{{\color{darkblue}7}}}{1}
     {8}{{{\color{darkblue}8}}}{1}
     {9}{{{\color{darkblue}9}}}{1}
     {:}{{{\color{darkblue}{:}}}}{1}
     {,}{{{\color{darkblue}{,}}}}{1}
     {\{}{{{\color{darkblue}{\{}}}}{1}
     {\}}{{{\color{darkblue}{\}}}}}{1}
     {[}{{{\color{darkblue}{[}}}}{1}
     {]}{{{\color{darkblue}{]}}}}{1},
}

\newlength{\upBranch} % shift up the text  lines <<<<
\newlength{\tolineSpace} % blank space bellow text  lines  <<<
\usepackage{xpatch} % needed <<<<<<<<

\makeatletter

\xpatchcmd{\dirtree} % root
{\vbox{\@nameuse{DT@body@1}}}
{\raisebox{-\tolineSpace}{\vbox{\@nameuse{DT@body@1}}}}
{}{}    

\xpatchcmd{\dirtree} % below space
{\advance\dimen\z@ by-\@nameuse{DT@lastlevel@\the\DT@countiv}\relax}
{\advance\dimen\z@ by-\tolineSpace \advance\dimen\z@ by-\@nameuse{DT@lastlevel@\the\DT@countiv}\relax}
{}{}
    
\xpatchcmd{\dirtree}% shift up the text  lines
{\kern\DT@sep\box\z@\endgraf}
{\kern\DT@sep\raisebox{-\upBranch}{\box\z@}\endgraf}
{}{}    

\makeatother

\jyear{2021}%

\theoremstyle{thmstyleone}%
\theoremstyle{thmstyletwo}%

\theoremstyle{thmstylethree}%

\newcommand{\smallerfontsize}{%
  \fontsize{9pt}{10pt}\selectfont
}

\begin{document}

\title[Hierarchical Research Fields Classification]{Hierarchical Classification of Research Fields in the ``Web of Science" Using Deep Learning}


\author*[1,2]{\fnm{Susie Xi} \sur{Rao}}\email{srao@ethz.ch}
% orcid: https://orcid.org/0000-0003-2379-1506

\author[1]{\fnm{Peter H.} \sur{Egger}}\email{pegger@ethz.ch}
% orcid: https://orcid.org/0000-0002-0546-1207

\author[2]{\fnm{Ce} \sur{Zhang}}\email{ce.zhang@inf.ethz.ch}
% orcid: https://orcid.org/0000-0002-8105-7505

\affil*[1]{\orgdiv{Department of Management, Technology, and Economics}, \orgname{ETH Zurich}, \orgaddress{\street{Leonhardstrasse 21}, \city{Zurich}, \postcode{8092}, \state{Zurich}, \country{Switzerland}}}

\affil[2]{\orgdiv{Department of Computer Science}, \orgname{ETH Zurich}, \orgaddress{\street{Stampfenbachstrasse 114}, \city{Zurich}, \postcode{8092}, \state{Zurich}, \country{Switzerland}}}

%\affil[3]{\orgdiv{Department}, \orgname{Organization}, \orgaddress{\street{Street}, \city{City}, \postcode{610101}, \state{State}, \country{Country}}}


\abstract{\smallerfontsize This paper presents a hierarchical classification system that automatically categorizes a scholarly publication using its abstract into a three-tier hierarchical label set (\textit{discipline}, \textit{field},  \textit{subfield}) in a multi-class setting. This system enables a holistic categorization of research activities in the mentioned hierarchy in terms of knowledge production through articles and impact through citations, permitting those activities to fall into multiple categories. The classification system distinguishes 44 disciplines, 718 fields and 1,485 subfields among 160 million abstract snippets in Microsoft Academic Graph (version 2018-05-17). We used batch training in a modularized and distributed fashion to address and allow for interdisciplinary and interfield classifications in single-label and multi-label settings. In total, we have conducted 3,140 experiments in all considered models (Convolutional Neural Networks, Recurrent Neural Networks, Transformers). The classification accuracy is $\gt$90\% in 77.13\% and 78.19\% of the single-label and multi-label classifications, respectively. We examine the advantages of our classification by its ability to better align research texts and output with disciplines, to adequately classify them in an automated way, and to capture the degree of interdisciplinarity. The proposed system (a set of pre-trained models) can serve as a backbone to an interactive system for indexing scientific publications in the future.}


%, O3}
% C38	Classification Methods • Cluster Analysis • Principal Components • Factor Models
% O3	Innovation • Research and Development • Technological Change • Intellectual Property Rights
%, 62R07, 68-04}
% 68T50: NLP
% 62R07: Statistical aspects of big data and data science {For computer science aspects, see 68T09; for information-theoretic aspects, see 94A16}
% 68-04: Software, source code, etc. for problems pertaining to computer science

\maketitle

\section{Introduction} \label{sec:intro}

% We need a hierarchical clf system.
For many purposes in academic life and beyond, a hierarchical classification \citep{bransford1999people, tsien2007memory} of academic output into disciplines, fields, and subfields appears desirable, if not mandatory. Let us provide three examples to illustrate the need.
% \footnote{As hierarchical structures are quite natural for human brains which utilize hierarchical structures to store/recognize information \citep{bransford1999people, tsien2007memory}. A hierarchical structure is helpful for users to combine the new information they receive with their prior knowledge.}  

\begin{enumerate}
    \item A young, talented high-school graduate wanted to choose a discipline and later a field and subfield of study, as well as the associated top institutions in terms of faculty for their focus according to the dynamics in the (sub-)field in terms of output and citations, in order to optimize their career prospect.
    \item A scientific funding institution wanted to determine the relative degree of interdisciplinarity in a field of study to judge applications in this regard.
    \item A university-level tenure committee wanted to objectively determine the top scholars of the same cohort in terms of their output and impact in the same subfield as a given candidate.
\end{enumerate}

% The current state of clf systems.
Each of these interests requires a clear delineation between disciplines, fields, and subfields across an array of academic domains of interest and, hence, a categorization of academic work and interests in horizontal (across disciplines; across fields within a discipline; across subfields within a field of study) and vertical terms (in discipline, field, and subfield of study).\footnote{For clarification, we hereby define our terminology in describing the hierarchy of fields in this paper. A \textit{discipline} is defined as an academic discipline, known as a branch of knowledge. Beneath which, its subfields are called \textit{fields} in this paper. The subfields under a field are called \textit{subfields} here. We have a three-level hierarchy defined in our paper to describe the structure of academic fields: \textit{discipline-field-subfield}. The specification of fields helps us to specify the analysis on different levels as we will see later. The term \textit{field} is used as a general term to refer to  a \textit{discipline} or a \textit{field} or a \textit{subfield}. } A few disciplines such as computer science, economics, mathematics, and physics have established widely agreed vertical systems within their discipline. However, a system with comparable granularity that encompasses most academic disciplines is missing. This void poses problems for a comparison of, for example, the 'narrowness' of fields in different disciplines, the degree of interdisciplinarity or impact breadth of work, or the relative performance of scholars with a similar focus of interest.

% related work: introduction about general meta-science approaches, and how our work contributes to this research domain
This paper contributes to the literature on meta-science -- the science of science -- which roots in scientometrics, bibliometrics, and informetrics. Specifically, it relates to efforts which focus on a comparison of scholars or academic institutions in specific disciplines and fields, on the measurement and outputs reated to interdisciplinary endeavors, on the dissimilarity or similarity of research bases, scholarly inputs and outputs of disciplines and fields, etc. 

% Motivation: We need large amount of data and a good clf system is needed.
For the aforementioned lines of interest, we need a unified classification system across disciplines and a large amount of data with good coverage of disciplines. Regarding the classification system, we have taken a primarily supervised machine learning (ML) approach. Regarding data, Microsoft Academic Graph (MAG) represents a particularly useful database for these approaches as it contains abstracts and other attributes on a large amount of scientific output, and the quality of the data it covers has improved significantly over the years \citep{sinha2015overview, shen2018web, MAGreview2019}.

%\footnote{MAG provides good coverage and is comparable to many academic databases, c.f.~\cite{martin2021google} for a small quantitative analysis of the coverage in English citations. In \cite{martin2021google}, they looked at six data sources (Microsoft Academic, Dimensions, the OpenCitations Index of CrossRef open DOI-to-DOI citations (COCI), Web of Science Core Collection (WoS), Scopus, or Google Scholar) and investigated 3,073,351 citations found to 2,515 English-language highly-cited documents published in 2006 from 252 subject categories based on the subject fields listed on Google Classic Papers (2006) (c.f.~the list \href{here}{https://scholar.google.com/citations?view_op=list_classic_articles&hl=en&by=2006}), expanding and updating the largest previous study \citep{birkle2020web}. One of the key advantages of MAG is that it exists as an open-sourced, and relatively well organized data dump (downloadable via Azure Data Share Services until Dec.~2021). See archived instructions for data access under \url{https://learn.microsoft.com/en-us/academic-services/graph/get-started-receive-data} (last accessed: Nov 18, 2022).}

With the supervised approach, the mappings of abstracts and keywords from MAG to the targeted classification system are informed (i.e., the algorithms are trained) by existing discipline classifications using cross-disciplinary information such as ``List of academic fields" from \cite{wikidis} and discipline-specific classifications, such as JEL \citep{jel} in economics, ACM \citep{acm} in computer science and PACS \citep{pacs} in physics and astronomy.

The main purpose of this paper is to deliver a system which acknowledges the organization of academic work horizontally between disciplines and vertically into discipline, field, and subfield in order to help answering questions of the aforementioned type. As indicated above, this approach faces the following challenges. Specifically, we need to delineate the boundaries between disciplines (e.g., mathematics, economics, engineering, computer science) as well as of fields and subfields within disciplines. In this regard, we decided to establish an ontology that has approximately even granularity across disciplines, fields, and subfields. A key goal of this ``global'' classification system is to enable normalized and unnormalized impact analyses within and across disciplines, while permitting a multi-label (interdisciplinary, interfield, intersubfield) classification of academic output based on abstracts and keywords. Finally, a prerequisite for such a system is that it can be modified and extended on the basis of new data in a timely manner, while using extremely large amounts of data (such as abstracts and keywords).

% key contributions
Using data from MAG and supervised algorithms, the paper offers the following key contributions. First, we develop advanced tools to classify large amounts of text data into clusters. Such an analysis is key when one wishes to understand important features of the state or its change over time of a field of interest here in the academic publication space. Specifically, we permit the clusters to overlap, so that any type of output can principally be classified as being situated strictly within or between (in multiple) disciplines, fields, and subfields. Specifically, we conduct a large number of experiments in various state-of-the-art neural network architectures (Convolutional Neural Networks, Recurrent Neural Networks, Transformers) and evaluate extensively a set of performance metrics (accuracy, precision, recall) in 44 disciplines, 718 fields, and 1,485 subfields in both single-label and multi-label settings to proposed the envisaged classification scheme. We make the codebase of the classification system publicly available and accessible under \url{https://gitlab.ethz.ch/raox/science-clf}.

The remainder of the paper is organized as follows. We introduce the research methodologies and our goals in designing a hierarchical classification system in Section~\ref{sec:clf}. In Section~\ref{sec:data-source}, we first discuss the utilization of data sources (MAG for academic abstracts and keywords and other sources for existing discipline label systems). Section \ref{sec:input2clf} describes the challenges and need to generate high-quality training data by linking the data sources underlying our classification system.  In Section~\ref{sec:sys} we propose the design of a modularized hierarchical classification system in both \textit{single-label} and \textit{multi-label} settings. We present our experimental setups and evaluate their performance in single-label and multi-label settings in Section~\ref{sec:eval}. Finally, in Section \ref{sec:interdisciplinarity}, we report on interfieldness scores \textit{within} and \textit{across} all disciplines with field as the unit of analysis as an exemplary result derived from the classification system.

%Then we discuss the creation of our three-level ontology (\textit{discipline-field-subfield}), the creation of discipline-publication mappings, as well as the preprocessing of abstracts to speed up the computation. 

% \section{Motivation}\label{sec:motivation}
% Evergrowing up-to-date publications and hard to grasp/search
%We are overwhelmed by the information flow we receive daily, and also the by the academic publications rolling in daily. As human beings, we would never catch up reading at the pace in which new papers are published. To retrieve papers one could read to be well informed, we usually go to one of the academic database products, search based on keywords or some other criteria (\textit{h-index}, citation counts, popularity of authors, publication venue, institutions, year) and try to single out the publications we want to read. 

% R&R note: explain we need a hierarchical, multi-class/multi-label clf, why there are useful in the present context

\section{Hierarchical Multi-Class Classification}
\label{sec:clf}
In this paper, we present a modularized \textbf{\textit{hierarchical}} \textbf{\textit{multi-class}} classification system, which is capable of handling a large amount of text data as contained in MAG (see Section~\ref{sec:mag}) in multi-level label schemes (see Section~\ref{sec:3lclf}). In a nutshell, the proposed classification takes an abstract of a publication as input and outputs three labels which indicate at least one discipline (e.g., computer science), at least one field (e.g., information system), and at least one subfield (e.g., database) the publication belongs to. The system is \textbf{\textit{modular}}, because it can cope with training and inference in a \textit{discipline-field-subfield} structure, and it can take any state-of-the-art neural architecture as classifier.

A \textbf{\textit{hierarchical}} classification, as opposed to a single classifier, is needed for the following reasons. First, a single classifier for all categories at the deepest levels could be used in conjunction with preexisting hierarchy information for disciplines. However, on the one hand, due to class confusability \citep{10.5555/2627435.2638582, 10.1145/1089815.1089821, 10.1007/s10994-011-5272-5} and class imbalance \citep{LIU2009690, PADURARIU2019736}, ML-based text classifiers usually perform poorly and become costly to train as the number of classes increases. This is avoided in the present context when working with disciplines, fields within disciplines, and subfields within fields, rather than treating all subfields horizontally and simultaneously. Moreover, even with subfields at hand, one would not easily be able to associate those with disciplines, as many disciplines do not have widely acknowledged hierarchical intra-disciplinary categorization schemes. Second, for the purpose at hand, one will need to re-train the models from time to time to keep the classifications constantly up to date, because scholarly publications are streamed in timely, and in the future one might wish to incorporate a ``human-in-the-loop" approach. Hence, it is costly to have a single model for all labels, which requires re-training the whole system every time one updates the input. In contrast, having several models in a hierarchy allows one to selectively train only those models that require an update. For instance, this means in our case one could re-train the model for a single discipline only with a new dump of discipline-specific publications delivered. 
% However, for search routines to produce high-quality output, and for questions \textcolor{blue}{we want to answer}, searching within a hierarchical structure is desirable if not a must.

%%%
Now, we discuss two settings in the \textbf{\textit{multi-class}} classification: (1) \textit{\textbf{single-label}}, where we assume that each piece of academic output (and abstract or paper) can only be assigned to one category; (2) \textit{\textbf{multi-label}}, where we assume binary relevance \citep{10.1007/978-3-540-24775-3_5, 10.1007/978-3-662-44851-9_28} of each category and the categories are independent of each other, with each piece of academic output being potentially assigned to multiple categories.\footnote{In future work, we would like to extend our current multi-label classification system to alow for a score of the discipline composition of an article. For example, one could then state that an article 30\% belongs to computer science and 70\% to economics. For this, one would have to remove the label independence assumption and construct a label powerset of the classes. In this regard, one could make use of data-driven modularity metrics, defining the class powersets of the labels (c.f.~\cite{szymanski2016data}).} 

The goal of the hierarchical multi-class system in the present context is two-fold: first, to provide a system that is modularized by disciplines, fields, and subfields, which enables efficient re-training of the models; second, to enable the system to perform both single-label and multi-label classifications in a multi-class setting. 

\section{Data Sources}\label{sec:data-source}

The present work uses the following data sources. 

\subsection{MAG}\label{sec:mag}
The MAG database provides us with abstracts of a large number of scientific publications (our \textit{input} in Section~\ref{sec:preprocess}). In this section, we discuss the merits and disadvantages of MAG as a main data source for the present purpose.

% R&R note: 
% emphasize MAG with own amendments
% subsection: start with the merits and disadvantages of MAG and 
    % merits: raw data and coverage

\subsubsection{Merits and Disadvantages of MAG}
\label{sec:mag-overview}
% general info on MAG
The systems and results we develop in this paper are based on the MAG snapshot (2018-05-17) obtained from MAS with around 160 million scholarly publications (i.e., excluding patent publications). The database includes all scholarly publications with their attributes such as title, authors, affiliations, venue, field of study (FOS), abstract, citation count, paper reference, etc. The tables in the database are linked through paper ID, author ID, affiliation ID, FOS ID, etc. To see the most recent MAG entity data scheme, including the name and type of each attribute, see \cite{magschema}. The way in which the database was created and improved over the years and how some attributes (such as FOS) were generated has been described in detail in \citet{shen2018web, sinha2015overview, MAGreview2019}. 

% merits
MAG offers the following merits to users. 
\begin{itemize}
    \item It provides keywords and even a loose hierarchy in the FOS scheme that are human-curated (by the author or publisher) or machine-generated.
    \item It provides a set of normalized tables that can be easily joined via ``Paper ID". Through these joins, meta-data of publications such as citations, authors, and their affiliations, venues, etc., are accessible. 
    \item It has been continuously updated until the end of 2021 and its successor OpenAlex \citep{priem2022openalex} has taken over most of its structure. 
\end{itemize}

% MAG con: hierarchical structure
A key question to ask and answer is why one would deem it insufficient to work with the existing FOS tags available in MAG. Essentially, three reasons come to mind, noting that MAG's approach had not been developed with the intent of organizing disciplines in a comparable granular structure.

First, MAG only distinguishes between 19 disciplines (top-level FOS) rather than the 44 disciplines commonly discerned. E.g., among the many commonly acknowledged disciplines, MAG does not consider linguistics or archaeology. We work with 44 disciplines and design our own classification system based on them (see Section~\ref{sec:3lclf}).

Second, the MAG FOS scheme had not been established with the perspective of a hierarchical classification scheme as the one targeted here. Rather, it is inherited from what certain academic publishers have provided, and the latter is augmented with unsupervised machine learning (ML; c.f.~``Field of Study Entity Discovery" in \cite{sinha2015overview}). As a consequence, the available FOS tags (topics) differ vastly in terms of their granularity, and they are not systematically comparable or aligned with the classifications in disciplines that consider widely acknowledged hierarchical structures for themselves such as computer science (ACM), economics (JEL), or physics and astronomy (PACS).\footnote{According to \cite{sinha2015overview}, the FOS tags were generated by seeding using existing keywords of good quality through name matching and some heuristic rules. This does not assure an acceptable or comparable level of granularity of FOS tags within and across disciplines. We have evaluated the topic hierarchy in the horizontal and vertical manner. For instance, consider the following subfields of computer science (CS) according to MAG: ``Natural language processing", ``Machine translation", ``BLEU" and ``Chinese Mandarin". These are put on the same horizontal level by MAG, which apparently is deficient. Ideally, ``BLEU" should be a level lower than ``Machine translation". Moreover, it is not intuitive to put ``Chinese Mandarin" as an FOS tag of the same hierarchical level as ``Machine translation". Looking at the discipline of economics, we observe similar problems. We also observe that the number of child levels (equivalent to fields in disciplines) varies largely for each topic. Therefore, it is impossible to construct a global taxonomy based on the FOS topic hierarchy provided by MAG without making use of the external discipline classifications.} 

Third, FOS tags often contradict author-declared classifications in disciplines where such declaration is customary and typically published with academic texts (e.g., in ACM, JEL, or PACS). 

All of the above calls for the development of a new hierarchical system of fine granularity, which subsequently would permit an analysis of research input or output across comparable categories.

% subsection: why is it better than the others (googlescholar, wos, scopus) as a subsection of MAG (comparison with others)
\subsubsection{Comparison of MAG with Other Databases of Academic Output} 
\label{sec:comparison-mag}

% We do not need these images.

% There are many academic databases there, MAG is the best alternative.
There are several academic databases that cover academic output across disciplines and could be used as the main source in our project, the most prominent being \textit{Web of Science} by \textit{Clarivate Analytics} \citep{wos}, \textit{Google Scholar} \citep{googlescholar}, \textit{Scopus} by \textit{Elsevier} \citep{scopus}, and \textit{Microsoft Academic Graph} (MAG) \citep{sinha2015overview, MAGreview2019, wang2020microsoft}. Despite the three shortcomings discussed in Section~\ref{sec:mag-overview}, we choose MAG as a data source for two reasons: (1) it provides good coverage of scientific publications in an open-sourced data dump (see \citep{martin2021google} for a small quantitative analysis of the coverage in English citations),\footnote{In \cite{martin2021google}, they looked at six data sources (Microsoft Academic, Dimensions, the OpenCitations Index of CrossRef Open DOI-to-DOI citations (COCI), Web of Science Core Collection (WoS), Scopus, or Google Scholar) and investigated 3,073,351 citations found in 2,515 English-language highly cited documents published in 2006 from 252 subject categories based on the subject fields listed in Google Classic Papers (2006) (c.f.~the list \href{here}{https://scholar.google.com/citations?view_op=list_classic_articles&hl=en&by=2006}), expanding and updating the largest previous comparative study \citep{birkle2020web}. Archived instructions for data access to MAG can be found in \url{https://learn.microsoft.com/en-us/academic-services/graph/get-started-receive-data} (last accessed: Nov 18, 2022). MAG is succeeded by OpenAlex.} and (2) it is constantly updated and has linkages of publications, authors and affiliations.

%% commented out: I am completely lost w sec 3.1.2! WE establish a hierarchy. Whether anyone provides it or not is irrelevant. We say in 3.1.1 that FOS is bad. There is no way to say that FOS is the (only) reason for why MAG is the choice….

\subsection{Discipline Classifications}\label{sec:3lclf}

% R&R note: We create a 3l label system by combining wikipedia and domain-specific ones.

In this subsection, we explain the need to create a hierarchical label system by combining ``List of academic fields" from Wikipedia and domain-specific classifications such as ACM for computer science or JEL for economics.

% R&R note: 
% subsection: start with the merits and disadvantages of existing discipline classifications and label systems  

\subsubsection{Merits and Disadvantages of Existing Discipline Classifications and Label Systems}

We have evaluated existing discipline/field classifications published nationally and internationally. They come mainly from two sources: research funding institutions and Wikipedia.

We present here a list of classifications from various major research funding institutions we examined in May 2018: 
\begin{itemize}
    \item \cite{dfg},
    \item \cite{jica}, 
    \item \cite{asrc},
    \item \cite{oecd},
    \item \cite{nsf}, 
    \item \cite{eec}.
\end{itemize}

However, not a single one of the classification schemes of the above institutions provides a comprehensive global hierarchical structure as targeted here (\textit{discipline-field-subfield}), and also the information contained in the various sources cannot be combined in a straightforward way. We therefore proceed by defining the hierarchy starting from the level of disciplines and gradually fill in a two-sublayer hierarchy based on within-discipline classification schemes. 

``List of academic fields" from Wikipedia turns out to be the most comprehensive classification with good coverage of disciplines and a well-organized hierarchy. We use the version \citep{wikidis} published in May 2018, consistent with the timeframe of the MAG data dump used here. In total, the Wikipedia scheme covers 55 disciplines in the entire hierarchy.  
% \footnote{\url{https://en.wikipedia.org/w/index.php?title=List_of_academic_fields&oldid=841131306} (last accessed: Feb 29, 2020).}  

%R&R note, to emphasize: Wikipedia's coverage, subdiscipline, some disciplines do not come with a 3l level model

\subsubsection{Establishing Our Three-Level Label Hierarchy}

% We create our own 3l clf scheme by xxxx.
In this section, we discuss the steps to create our global classification scheme. It is done by combining the 44 disciplines in the Wikipedia ``List of academic fields" and discipline-specific classifications such as JEL and ACM. Our three-level hierarchy of \textbf{\textit{labels}} is used in Section~\ref{sec:dis-pub-map} to generate a high-quality training set for our own classification system.

\begin{figure*}
\DTsetlength{1em}{1em}{0.5em}{0.3pt}{1pt}       

% The first argument (1em) sets the minimum width of the tree nodes.
% The second argument (0.5em) sets the horizontal space between sibling nodes.
% The third argument (0.5em) sets the vertical space between parent and child nodes.
% The fourth argument (0.1pt) sets the thickness of the tree lines.
% The fifth argument (0.8pt) sets the thickness of the connection lines between parent and child nodes.
% \setlength{\DTbaselineskip}{5pt}:

% This command sets the baseline skip, which determines the vertical space between lines of text within the tree nodes. In this case, it's set to 5pt, indicating a smaller vertical space between lines.
% \renewcommand{\DTstyle}{\rmfamily\tiny}:

% This command redefines the style of the tree nodes. In this case, it sets the font family to \rmfamily (Roman) and the font size to \tiny (very small).

%\setlength{\DTbaselineskip}{12pt}  % minimum size for \footnotesize 
%\renewcommand{\DTstyle}{\rmfamily\footnotesize} 
    
%\setlength{\DTbaselineskip}{16pt}  %minimum size for \normalsize
%\renewcommand{\DTstyle}{\rmfamily\normalsize} 

\setlength{\DTbaselineskip}{5pt} %minimum size for  \large
\renewcommand{\DTstyle}{\rmfamily\footnotesize}    %

\dirtree{%
.1 \textbf{ROOT}.
.2	Humanities.
.3	Anthropology.
.4	Archaeology (*).
.3	History (*).
.3	Linguistics and languages (*).
.3	Philosophy (*).
.3	Religion (*).
.3	The arts.
.4	Literature (*).
.4	Performing arts (*).
.4	Visual arts (*).
.2	Social sciences.
.3	Economics (*).
.3	Geography (*).
.3	Interdisciplinary studies.
.4	Area studies (*).
.4	Ethnic and cultural studies (*).
.4	Gender and sexuality studies (*).
.4	Organizational  (*).
.3	Political science (*).
.3	Psychology (*).
.3	Sociology (*).
.2	Natural sciences.
.3	Biology (*).
.3	Chemistry (*).
.3	Earth sciences (*).
.3	Physics (*).
.3	Space sciences (*).
.2	Formal sciences.
.3	Computer  (*).
.3	Logic (*).
.3	Mathematics.
.4	Pure mathematics (*).
.4	Applied mathematics (*).
.5	Statistics.
.3	Systems science (*).
.2	Professions and applied sciences.
.3	Agriculture (*).
.3	Architecture and design (*).
.3	Business (*).
.3	Divinity (*).
.3	Education (*).
.3	Engineering and technology (*).
.3	Environmental studies and forestry (*).
.3	Family and consumer science (*).
.3	Human physical performance and recreation (*).
.3	Journalism, media studies and communication (*).
.3	Law (*).
.3	Library and museum studies (*).
.3	Medicine (*).
.3	Military sciences (*).
.3	Public administration.
.4	Public policy (*).
.3	Social work (*).
.3	Transportation (*).
}
\caption{Discipline hierarchy of the Wikipedia taxonomy. Note that we only have to classify the leaf nodes, which leaves us with 44 disciplines (marked with (*)).}
\label{fig:wiki}
\end{figure*}

\paragraph{\textbf{Step 1: Pruning the Wikipedia Hierarchy.}}
We list the discipline hierarchy of the Wikipedia taxonomy in Figure~\ref{fig:wiki}. Note that we only have to classify the leaf nodes, which leaves us with 44 disciplines (marked with (*) in Figure~\ref{fig:wiki}). For instance, knowing the classifications of ``Literature", ``Performing arts" and ``Visual arts" gives us the whole hierarchy of their parent discipline ``The arts", so we do not need to run a classification for the parent discipline ``The arts". 

\paragraph{\textbf{Step 2: Merging Discipline-Specific Classifications with the Pruned Wikipedia Hierarchy.}}
% (to reorg) This describes our usage of existing ACM/JEL taxonomies
What we need next is a vertical (or hierarchical) classification scheme for each of the 44 disciplines. Here, we make use of existing discipline classifications, such as the ACM classification for computer science \citep{acm}, the JEL classification for economics \citep{jel}, the PACS classification for physics \citep{pacs}, and the MSC classification for mathematics~\citep{msc}. 

Note that for the classification scheme of mathematics, we carefully compare MSC vs.~the classification of mathematics in the Wikipedia taxonomy (where ``Mathematics" $\rightarrow$ ``Pure mathematics" $+$ ``Applied mathematics"). Instead of MSC, we decided to use ``Pure mathematics" and ``Applied mathematics" as two fields for mathematics. This is because of the overrepresentation of mathematics if we use MSC as a discipline classification.

%\footnote{\url{ftp://cran.r-project.org/pub/R/web/classifications/MSC.html} (last accessed: May 23, 2018). } 

For those disciplines that do not have a pre-defined classification, Wikipedia serves as the source. For instance, for the discipline ``Linguistics and languages", we resort to the corresponding linked page \citep{wikill} in the ``List of academic fields". 

% \footnote{\url{https://publishing.aip.org/publishing/pacs/pacs-2010-regular-edition} (last accessed: May 23, 2018).} 

\paragraph{\textbf{Step 3: Create a \textit{Discipline-Field-Subfield} Label Structure for All Disciplines.}}
It is worth noting that not every discipline classification has a three-level structure in Wikipedia or its own classification system. We decide for each discipline to feature such a three-level structure, because it is common to have one for the majority of disciplines (c.f.~``List of academic fields" from Wikipedia \citep{wikidis}). We take the 44 disciplines as the \textit{discipline} level, the first level tags in each discipline classification as the \textit{field} level and the second level tags in each discipline classification as the \textit{subfield} level. Hence, our classification scheme is defined as a \textit{discipline-field-subfield} structure (see the illustration of JEL classifications in Figure~\ref{fig:dis-pub-map-econ}).

\section{\textbf{From Abstracts and Labels to A Classification}}
\label{sec:input2clf}
The following subsections describe how we connect the data sources, which include both abstracts (Section \ref{sec:mag}) and labels (Section \ref{sec:3lclf}), with a classification system (Section \ref{sec:3lsys}). We started by using the existence of FOS tags to link the labels with the abstracts, but found that only 51.3\% of the abstracts could be assigned to labels (see Section \ref{sec:matching-abstract-label}). To improve on the latter, we experimented with supervised and unsupervised topic models (Appendix~\ref{app:topic-model}), as well as simpler supervised models such as SVM (Appendix~\ref{app:h-svm}). But those approaches did not perform well in disambiguating disciplines. Ultimately, we created a modularized three-level hierarchical classification system (Section~\ref{sec:sys}) that supports a subsequent analysis, for which we provide an example related to research interfieldness and interdisciplinarity (Section \ref{sec:interdisciplinarity}). The modular approach enables us to break down the process into manageable parts and better understand the connections between the data sources and the classification system. 

%%%%%%%%%%

% R&R: add in Section 3 the % we map already with the new clf
% why we need a new clf system, because with the above label system we can only cover xxx % and we want a modularized system so that we can also inspect the quality. We also want to study interdisciplinarity with this system.
\subsection{Linking Abstracts and Labels} \label{sec:matching-abstract-label}

\begin{figure}
  \centering
  \includegraphics[width=\linewidth]{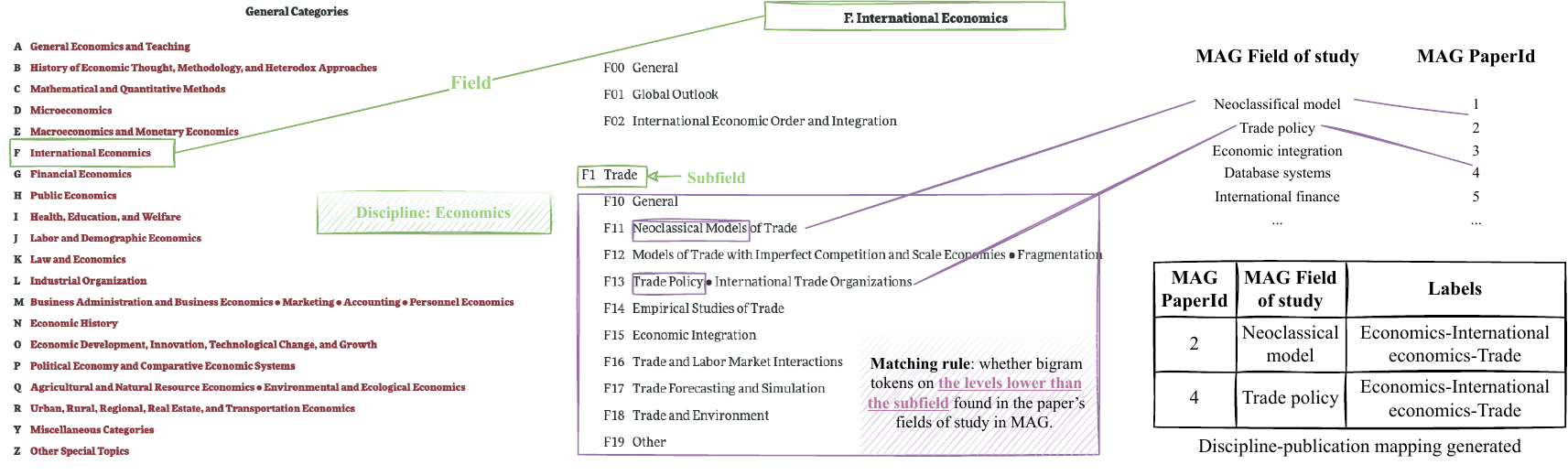}
  \caption{Discipline (JEL) publication mapping using FOS tags from MAG.}
  \label{fig:dis-pub-map-econ}
\end{figure}

To create a discipline-publication mapping, we start linking the abstracts and labels. Here, we describe how we automatically annotate a publication with a set of \textit{discipline-field-subfield} labels. 
\begin{itemize}
    \item For the disciplines of ``Computer science", ``Economics" and ``Physics", where a discipline-specific annotation exists (ACM, JEL, and PACS, respectively), we run name-matching based on the existence of FOS tags in the levels lower than the third level (e.g., levels 4-6 for JEL and ACM). We do this only for the subfield level. 
    \item For the disciplines without a pre-defined taxonomy, we use the Wikipedia taxonomy (e.g., ``Linguistics and languages" in \cite{wikill}) and find the matches of the FOS tags in the text description provided by Wikipedia (e.g., \cite{wikill_def} for ``Linguistics and languages"). Likewise, we do this only for the subfield level. 
\end{itemize}
Specifically, we first distinguish between singleton and non-singleton FOS tags. Non-singleton FOS tags are matched based on their presence in the text describing the subfield in the Wikipedia taxonomy (tag $\langle wikitext \rangle$ \href{here}{https://gitlab.ethz.ch/raox/science-clf/-/blob/main/Wikipedia_hierarchy/WIKIXML_version20180723_full_topicwords.xml}).
For singleton FOS tags, we match based on their appearance in topic nouns (tag $\langle TopicNouns \rangle$ \href{here}{https://gitlab.ethz.ch/raox/science-clf/-/blob/main/Wikipedia_hierarchy/WIKIXML_version20180723_full_topicwords.xml}), where the topic nouns are extracted using LDA topic modeling \citep{blei2003latent} and setting the topic to 1 for the text in each subfield. If there is a match between a subfield and FOS tags in one publication, we annotate this publication with the corresponding set of \textit{discipline-field-subfield} labels. Figure~\ref{fig:dis-pub-map-econ} shows how the matching is performed following the JEL classification scheme in the discipline ``Economics". The matching rate, which states how many publications could be linked to a set of \textit{discipline-field-subfield} labels, was only 51.3\% (or 89,746,934 out of 174,910,379 individual research output identifiers in MAG).

The outcome mentioned above prompts us to consider methods for assigning labels to all papers that have not yet been matched. The challenges involved in this task include (1) the need for a rapid and large-scale approach and (2) the importance of quality, which requires a robust model with high performance.

\subsection{\textbf{Related Work on Abstract-to-Discipline Classification and Baseline Models}}

% related work in field classification systems
\cite{kowsari2017hdltex} have proposed a two-level hierarchical classification system to classify scientific fields in a \textit{single-label} setting. They have investigated only seven disciplines (``Biochemistry", ``Civil engineering", ``Computer science", ``Electrical engineering", ``Medical sciences", ``Mechanical engineering", ``Psychology") with a small web-crawled dataset (WOS-46985) as a proof of concept. Their codebase is publicly available, but it suffers from scalability issues in both data loading and embedding computation, and it does not support parallel training. We adopt the same concept of hierarchical text classification but propose a new usage of data sources (MAG, Wikipedia classification, and in-domain classifications such as ACM, JEL, and PACS), and build a modularized pipeline scalable to the one of the largest body of academic publications. 

We have conducted the case studies illustrated in Appendix~\ref{app:case-study-econ-cs} of the disciplines ``Computer science" and ``Economics" from MAG and on a benchmarking dataset WOS-46985 published in \cite{kowsari2017hdltex}. As baseline models, we use topic modeling and hierarchical support vector machines (SVM). We also evaluated deep learning-based models described in \cite{kowsari2017hdltex}. From this effort, we concluded that the deep learning-based models developed for hierarchical classification as described in the Appendix~\ref{app:2lsys} suited our purpose the best. We designed our large-scale three-level classification system, with an efficient data loader built on top of the Lightning Memory-Mapped Database (LMDB) \citep{lmdb} and the modularized model training and inference described in Section~\ref{sec:3lsys}.

% R&R: the above section discusses why and we now show how.
% we mention single-label and multi-label, also modularized, one can choose different NN architectures. 

% then the result section and the preliminary result table.
\section{Our Proposed Classification System}\label{sec:sys}
We now present our modularized three-level hierarchical classification system by introducing the system design -- modularizable neural architecture as classifiers (Section \ref{sec:3lsys}), the single-label and multi-label settings (Section \ref{sec:single-multi-setting}) in the classification, and the preprocessing steps of our system input -- abstracts and labels (Section \ref{sec:preprocess}).

\subsection{Three-Level Classification}\label{sec:3lsys}

\begin{sidewaysfigure}
%\begin{figure*}
  \centering
  \includegraphics[width=\linewidth]{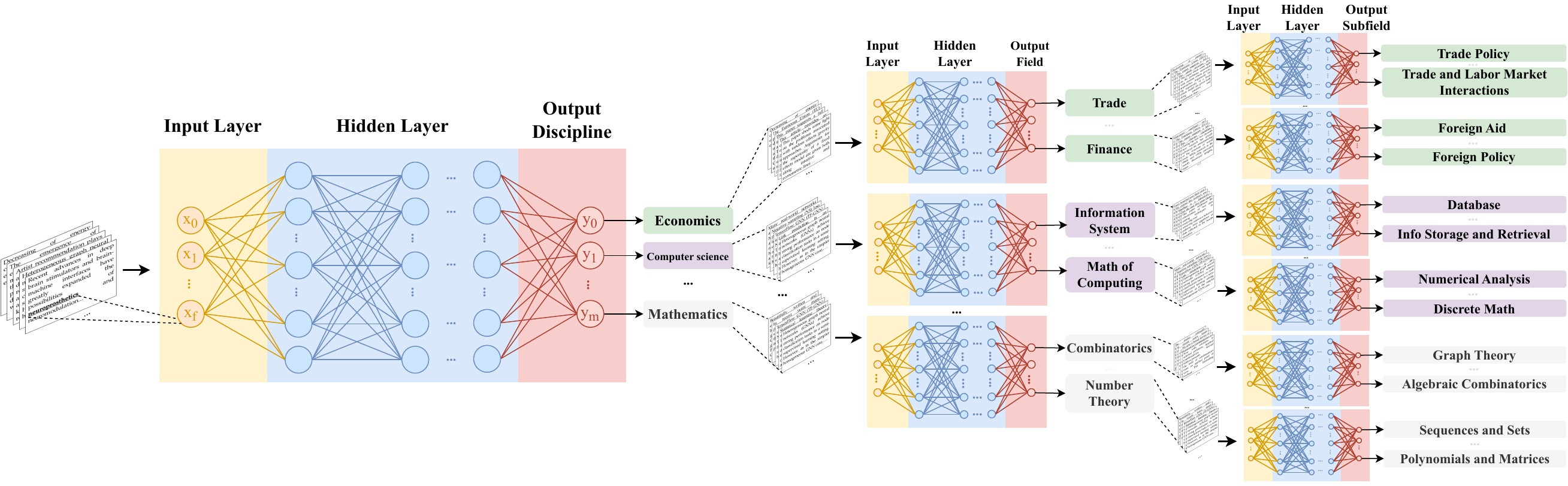}
  \caption{Three-level hierarchical classification system.}
  \label{fig:3lsys}
%\end{figure*}
\end{sidewaysfigure}

% System components
We design a three-level classification system as depicted in Figure~\ref{fig:3lsys}. The input of the whole system is an abstract of an article in the training corpus. And the output for each publication is a triplet of labels (\textit{discipline, field, subfield}). The system can be trained in a distributed way and has the capability to handle large datasets thanks to the preprocessing in Section~\ref{sec:abstracts}. 

\paragraph{\textbf{System Components.}}
We design the system in a modular fashion; this means that users can easily adapt the system with newer and fancier deep learning models (e.g., Transformers \citep{devlin-etal-2019-bert}) and with any hierarchical taxonomy that has a similar structure to ours. There are three components in our classification system. The first component (L1) performs classification in \textit{disciplines}, the second component (L2) in \textit{ fields}, and the third (L3) in \textit{subfields}. In each component, we have implemented four architecture choices, as discussed later in this section, feedforward deep neural networks (DNN), recurrent neural networks (RNN) using gated recurrent units (GRU), convolutional neural networks (CNN), and Transformers. 

The hierarchical system can help to determine the assignment of a research abstract to disciplines, fields, and subfields. We denote $p$ a publication, $D$ a discipline, $F_i$ a field in $D$, $F_{ij}$ a subfield in $F_i$. We obtain the unconditional probability $P(p \in D)$ from the first component of the system that classifies the disciplines and $ P(p \in F_i \mid p \in D) $ from the second component that classifies the fields. Similarly, one can compute the composition to a finer granularity by getting the $P(p \in F_{ij} \mid p \in F_i, p \in D)$ from the third component that classifies the subfields.

The second component (L2) of the classification system consists of a neural network (say DNN) trained for the domain output in the first component (L1). The input in a neural network in the second component (e.g., DNN) is connected to the output of the first level. 
For example, if the output of the first component is labeled ``Computer science" ($D$) then the DNN in the subsequent component to predict fields $F_i$ in $D$ is trained only with all computer science publications. Therefore, while the DNN in the first component is trained with all publications of all disciplines, each DNN in the second component is trained only with the publications for the specified discipline $D$. This applies to the third component (L3), e.g., only the abstracts that are classified to belong to ``Information system" ($F_i$) are then fed into the neural network that classifies the subfields under ``Information system" in the third component.

%Susie: single-label and multi-label (in each we have a modular structure and transformer is one of the best-performing models)
%Susie: move up the transformer

% note: do not use [ht!] flag in sidewaysfigure, otherwise the figure appears only at the end of the paper

\paragraph{\textbf{Modularized Neural Classifiers.}}

One advantage of a modularized hierarchical model is that, at a certain level, a poorly performing model (like CNN) could be easily replaced by a stronger model (like Transformers), without needing to change other submodels or data input. We now list the neural classifiers implemented in our system, DNN, RNN using GRU, CNN, and Transformers. Additional neural architectures can be easily integrated into the modularized system.

DNNs are well suited for text classification tasks because they are capable of learning complex non-linear relationships between input features and output labels. The input features could be words or word embeddings, and the output labels could be class labels, such as topic labels. DNNs can handle inputs of variable length by using padding or truncation techniques. However, they do not have any memory mechanism to handle sequential data, making them unsuitable for long sequential text data.

RNNs are designed to handle (long) sequential data and can capture temporal dependencies in input data. In the context of text classification, RNNs can take a sequence of words as input and use their memory to capture the context and meaning of each word in the entire sequence. This makes RNNs ideal for tasks with strong context dependency. RNNs can be trained using backpropagation through time, which updates the weights of the network based on the error at each time step. We use GRUs (instead of LSTMs) for computational efficiency or simplicity in our design.

CNNs are primarily designed for image classification, but have also become widely used for text classification. With the latter, CNNs can be used to extract local features from text data by treating them as two-dimensional signals. This is achieved by using two-dimensional convolutions over the sequence of words, which allow the network to capture patterns and relationships between adjacent words in the input. This makes CNNs ideal for tasks such as text categorization, topic classification, and sentiment analysis. In addition, CNNs are computationally efficient and can be trained on large datasets.

Transformers use self-attention mechanisms to model contextualized relationships between words, build a stack of encoders (``transformer blocks''), and learn the word representations in ``weights''. These weights then determine the importance of words/sentences for further processing. Bidirectional Encoder Representations from Transformers (BERT) are pre-trained\footnote{BERT has been trained with English Wikipedia of approximately 2.5 billion words and BooksCorpus of approximately 800 million words \citep{7410368}.} Transformer models on tasks like the masked-sentence task and next-sentence prediction. BERT outperforms baselines on many other tasks such as Q\&A \citep{devlin-etal-2019-bert}. It is ``bi-directional'' in the sense that words before and after the target word-to-predict are considered. For transfer learning where the embedding representations could be obtained from pre-trained BERT models, one does not need to retrain the in-domain models from scratch but only need to freeze most of the layers and fine-tune the few last layers. In fine-tuning, we jointly train them with newly added layers, such as dense layers, for downstream specific classification. 

BERT models are composed of two parts, the pre-processing and the encoder fine-tuning parts. The pre-processing encoder generates three representative embeddings given an input text into: (1) ``token'' embeddings based on the present tokens, (2) ``segment'' embeddings based on the sentence or phrase a particular token belongs to, and (3) ``positional'' embeddings based on the token position. Similar to the maximum sequence length imposed on the tokenization for DNN/CNN/RNN, BERT has the limit of 512 tokens of each input text. The encoder part uses the BERT pre-trained model, transforms an input text into embeddings via the pre-processing steps we discussed above, then passes the embeddings to the transformer blocks, and finally puts a dense layer as we discussed in the fine-tuning procedure.

%%%%%%
\subsection{Single-Label and Multi-Label Settings}
\label{sec:single-multi-setting}

In single-label settings, each piece of academic output is assigned to only one category, while in multi-label settings, each output can be assigned to multiple categories, with binary relevance and independent categories. These concepts have been discussed in related work such as \citet{10.1007/978-3-540-24775-3_5, 10.1007/978-3-662-44851-9_28}.

\paragraph{\textbf{Necessity of Multi-Label Classification and Its Assumptions.}}
In the context of interdisciplinarity (Section \ref{sec:interdisciplinarity}), if we only look at the single-label setting in a \textit{multi-class} classification, where a single input is associated with \textit{exactly one} of the many potential classes, it is a strong restriction. When we consider the \textit{multi-label} setting, where a single input is allowed to be associated with \textit{one or more} classes. This way, cross-disciplinary inputs such as ``Biochemistry" -- which combines ``Biology" and ``Chemistry" -- are considered to have two labels. 

As the simplest approach to start with, multi-label classification can be viewed as multiple binary classification problems under the assumption of label independence. That is, given one input, the relevance of one label does not depend on the relevance of other labels. Another assumption to make for multi-label classification is the train-/test-split. In the single-label setting, the split can be done by stratified sampling the label using its original distribution, since each input belongs to precisely one label. In the multi-label setting, however, since each input can belong to more than one label, simply performing ``stratified'' sampling is no longer trivial. Moreover, using the label powersets to perform stratified sampling is highly impractical: there are many possible combinations of labels up to $2^n$ for $n$ classes, so we might end up having very few samples belonging to some label set, which renders stratified sampling useless. The current implementation uses the same train-/test-split as in the single-label setting.\footnote{We are aware of that this setting can lead to the same sample might be assigned to both the train and the test sets from different labels it is associated with. One remedy for this problem is to use \textit{iterative stratification}, which allows setting up a $k$-fold cross-validation such that the distribution of relevant and irrelevant samples of each label is normalized \citep{sechidis2011stratification, pmlr-v74-szymanski17a}. Iterative stratification is implemented in the \texttt{scikit-multilearn} library for Python \citep{scikit-multilearn}. However, due to the lack of maintenance and documentation, we do not use this library at this stage and simply adopt stratified sampling as we do in the single-label setting.} 

\paragraph{\textbf{Specifications of Multi-Label Settings.}}

We conduct the experiments under the binary relevance (BR) assumption because of its scalability. Also, the hierarchical arrangement of the labels to some extent incorporates the label independence assumption. BR requires a minimal change to our technical setup as opposed to the single-label setup.

\begin{itemize}
    \item The input format does not need to change since the labels are already inputted as multi-hot encoding.
    \item The loss function must change from categorical cross entropy to binary cross entropy\footnote{Each pair of classes gets compared and their binary cross-entropy loss computed; the sum of losses for all the pairs is optimized.} following the BR approach, which is also found to be an effective loss function \citep{10.1007/978-3-662-44851-9_28}.
    \item The activation function of the final dense layer becomes \textit{sigmoid} instead of \textit{softmax}, following the loss of binary cross-entropy.
\end{itemize}
% \footnote{Sigmoid: $\mathbb{R} \to (0, 1)$, $x \mapsto \frac{1}{1 + \exp(x)}$.}

After changing to \textit{sigmoid} activation, the output array is thus no longer a probability array (i.e., its entries do not sum up to one). Each entry, which corresponds to the relevance of each label, remains between 0 and 1 (inclusive), which corresponds to the ``relevance'' of the input given one label. With this output format, it is also useful to set a threshold to treat the label as relevant: instead of cutting off at 0.5 as in normal \textit{sigmoid}, we might lower the cut-off  threshold, since the use case of the system is to explore the class membership.

As discussed above, we decide to perform stratified sampling on the label sets. This does not require any change to our codebase of the single-label setting.

Another choice we need to make is the performance metric, which we have used, namely \textit{categorical} accuracy, in the single-label experiments. The main difference between categorical accuracy and binary accuracy in a multi-label setting is that the latter exaggerates the performance when the ground-truth multi-hot label is sparse.\footnote{For example, consider the case where a classifier always predicts nothing, i.e., the predicted label is a zero array. When the ground truth label is sparse, for example, having only 2 out of 8 classes relevant, the \textit{binary} accuracy would be 6/8 = 75\%, although this classifier is completely uninformative. See \cite{binacc-tensorflow} on binary accuracy and \cite{catacc-tensorflow} on categorical accuracy for more details.} Therefore, we maintain the very conservative metric, categorical accuracy, along with precision and recall. Lowering the accuracy threshold from 0.5 to 0.3 may improve the performance; we will consider more sophisticated and customized metrics in future work.

% Preprocessing of Abstracts in MAG and our Labels
\subsection{Preprocessing of Input to Our Classification System}
\label{sec:preprocess}

Here we present the necessary preprocessing steps for the input (abstracts and label sets, see Section \ref{sec:data-source}) to our classification system. It should be noted that the same steps are used in both single-label and multi-label settings.

\subsubsection{Abstracts}\label{sec:abstracts}

We are in need of an efficient preprocessing pipeline for a system with a large number of training instances (984,722,678 abstracts in a multi-label setting). Each abstract is a text snippet of around 200 words, with variation across fields. 
\paragraph{From Raw Text to LMDB.}
Our model inputs -- the abstracts of the papers -- are provided in \texttt{PaperAbstractsInvertedIndex.txt} in the inverted index format in the MAG dataset. A dummy dataset is provided in Table \ref{table:mag-ii} for illustration purposes. We first decode all inverted abstracts in abstracts of normal reading order (the column ``\textit{Original abstract}"). We then tokenize each abstract into a sequence of its token IDs. We use \texttt{tensorflow.keras.layers.TextVectorization} for this.

We then store text vectors after tokenization in an LMDB instance per discipline, where batch generator IDs are keys, and the token sequences of their abstracts are values. This is to facilitate batch processing during training.\footnote{To ensure that the size of the training corpus does not become the bottleneck of training, because the existing preprocessing wrappers in e.g.,\textit{tensorflow} and \textit{keras} break down when computing the word index of the training corpus and then one hot encoding for each text snippet for a large training corpus. In our case studies discussed in Appendix~\ref{app:case-study-econ-cs}, more than 2 million training instances have already crashed the \texttt{fit\_on\_texts(texts)} method; see this post under \url{http://faroit.com/keras-docs/1.2.2/preprocessing/text/} (last accessed: Feb 29, 2020). }

\begin{table}[h!]
\centering
\caption{Dummy sample of paper abstract inverted index from MAG.}
\label{table:mag-ii}
\begin{tabular}{c|c|c|c}
\toprule
\textbf{Paper ID} & \textbf{IndexLength} & \textbf{InvertedIndex} & \textit{\textbf{Original abstract}} \\ \midrule
12               & 5                    & \begin{tabular}[c]{@{}c@{}}\{``I'': {[}0, 3{]}, \\ ``am'': {[}1, 4{]}, \\ ``who'': {[}2{]}\}\end{tabular} 
                                        & \textit{I am who I am}              \\ \midrule
1                & 3                    & \begin{tabular}[c]{@{}c@{}}\{``All'': {[}0, 2{]}, \\ ``in'': {[}1{]}\}\end{tabular}                                                                       & \textit{All in all}                 \\ \midrule
6                & ...                  & ...                    & \textit{...}  \\ \midrule
\end{tabular}

\end{table}

\paragraph{Efficient Data Loader with LMDB.}

We first select the top-$k$ frequent words from the ``bag of words" in the training set (say $k = 3000$) and pre-compute the representations with GloVe \citep{pennington-etal-2014-glove} or BERT that are static over all the models but fetched constantly in a key-value database (DB) that supports multithreading reads. After benchmarking many existing DB solutions (c.f. \cite{piriMT}), we select Lightning Memory-Mapped Database (LMDB) \citep{lmdb-doc} as our key-value store.

Making use of the word representations, we also pre-compute the representation for each abstract and store it in an LMDB instance. Each discipline of the 44 chosen has its own LMDB instance. This choice is a trade-off of data loading for different levels of models: typically, in the three-tier labels (\textit{discipline-field-subfield}), we most likely will update field-specific models or subfield specific models. The top level (aka a model to classify all 44 disciplines) will only be updated in case of a large update of the publication database.

Making data loading more efficient for all models gives our classification system a computational speedup and a capacity to handle an arbitrarily large number of training instances, which is expected as the scholarly publication space is growing rapidly. 
%%%

\subsubsection{Labels}\label{sec:labels}

Categorical labels are already provided after the discipline-publication mapping described in Section~\ref{sec:input2clf}, so preprocessing involves only one step: To encode categorical labels into category IDs readable by the classifiers. To incorporate the extension to both single-label and multi-label classifiers, we use multi-hot encoding using
\texttt{sklearn.preprocessing.MultiLabelBinarizer} for label encoding.

\section{Experiments and Evaluation}\label{sec:eval}

We perform experiments using the four neural network architectures (DNN, CNN, RNN, Transformer) described in Section \ref{sec:3lsys}. Note that due to the computational costs of Transformers, we only use them to improve the models that perform poorly with the basic neural architectures (DNN, CNN, RNN). We share the results of single-label and multi-label settings. 

The models are implemented in Python using the Keras library and are automatically tracked by the MLFlow library\footnote{MLFlow is an open-source platform for managing the end-to-end machine learning lifecycle \citep{10.1145/3399579.3399867, zaharia2018accelerating}. Its main functionality is the automatic tracking of different machine learning experiments and runs. In particular, it automatically records the training parameters and results as well as the trained models.} We use Keras \textit{Functional API}, which builds a directed acyclic graph (DAG) of layers to allow non-linear topology, such as shared layers and multiple inputs and outputs. We utilize the Distributed Learning module from the Tensorflow Distributed Learning API called \texttt{MirroredStrategy}, which supports synchronous training across multiple GPU workers,\footnote{One replica is created per GPU device, and each model variable is mirrored across all replicas and is kept in sync by applying identical updates. \textit{All-reduce algorithms} are used to communicate variable updates, in particular to aggregate gradients produced by different workers that work in different slices of input data in sync, using the NVIDIA Collective Communication Library (NCCL).} for fast training.

Categorical accuracy, precision, and recall are tracked. Other model settings can be found in the Appendix~\ref{app:training-setup}.

\subsection{Training and Testing Sets}
\label{sec:dis-pub-map}

The ultimate goal we want to achieve by linking the MAG abstracts and the discipline hierarchy (Section \ref{sec:input2clf}) is to create high-quality training data for our classification system. If there are matches between a publication abstract and a set of \textit{discipline-field-subfield} labels, we use this match in the training set; otherwise, a publication is put in the test set. Despite the abstract-label linkage only covering 51.3\% of all the papers in MAG, this approach is effective in automatically generating high-quality training instances for our classification system. The 40\% of the training set is used as the validation set and all the results we report below are on the validation set. The results are calculated using a generic list of the top vocabulary $k$ in the training corpus ($k=3000$).\footnote{We have experimented in \cite{piriMT} that the use of technical jargon from the abstracts can slightly improve the classification performance. However, it is not feasible to use technical jargon for all disciplines for the following reasons: (1) Not all papers have good FOS tags, as we see in the matching process in Section \ref{sec:dis-pub-map}. (2) It requires an additional step to extract quality words as discussed in \cite{rao2022keyword}. After careful evaluations of the existing methods of keyphrase extraction, we are currently using KeyBERT \citep{grootendorst2020keybert} to extract missing keyphrases for the MAG or OpenAlex data dump, which once completed can serve as the backbone of an upgraded version of the classification system.}

\subsection{Single-Label Experiments}
\label{sec:single-label-eval}

%%%%
% , as this minimizes the cost to generate a cross-44-discipline vocabulary list as we introduce the procedures in Appendix~\ref{app:kw-vocab}

%For the poor performing models (with an accuracy score less than 70\%, we can use a better vocabulary list to improve the quality as we have conducted in our ablation studies introduced in Appendices~\ref{app:vanilla} and \ref{app:kw-vocab}). 

\paragraph{CNN and RNN Results.}
In our small-scale ablation study described in \cite{piriMT} using the dataset described in Appendix \ref{app:case-study-econ-cs}, RNN and CNN significantly outperformed DNN. Therefore, we evaluate only the RNN and CNN models on the 44-discipline dataset. 
We performed in total 1,526 experiments: 2 architectures $\times$ 763 models (1 model for Level 0, 44 models for Level 1, 718 models for Level 2).  Due to the sheer number of experiments, the full result table is available in our code repository.\footnote{The results of single-label experiments are accessible under \url{https://gitlab.ethz.ch/raox/science-clf/-/blob/main/result_tables/single_label_clean_modelnotempty_export_v2_fixed.csv}.} 
%in the Appendix \ref{app:result-full-vanilla}. 
Here, we provide an executive summary of the results.

\begin{itemize}
    \item For most models, all architectures achieve good performance in all performance metrics. Specifically, in 77.13\%, 81.26\%, and 76.02\% of the models, their accuracy, precision, and recall reach 90\% or more regardless of the architecture, respectively.
    \item In terms of precision, CNN seems to perform best for most models, followed by RNN. Specifically, 54.95\% models get the best precision from CNN, and 45.05\% models from RNN. The mean precision scores for CNN and RNN in all models are 95.96\% and 93.70\%, respectively. The same conclusions can be drawn if we use accuracy and recall as performance metrics. 
    \item In the 165 models where the best accuracy of all architectures is less than 90\%, CNN also performs significantly better than RNN in all except 19 models. Looking at the best precision scores of all architectures, which are $\lt$ 90\%, we have 35 RNN models and 70 CNN models.
    %\item In the 1,189 models where the best accuracy is at least 90\%, the average precision is also at least 99.58\%, and the average recall is at least 97.81\%. (also deleted from code, not necessary to keep track of this, as we are only interested in the max accuracy below 0.9)

    \item For 4.7\% of all CNN/RNN models, their precision is extremely poor (less than 70\%) in any architecture. In these models, the mean accuracy is 50.87\%, and the mean recall is also poor with only 30.85\%. On average, there are 5.97 classes to predict in these models, with 1.35 million training instances on average.\footnote{Other statistics on the number of training instances are minimum 869, median 51,223, maximum 19,613,670. } Hence, we need to and should be able to further improve the performance of these models.
\end{itemize}

It is clearly beneficial to use CNN as the base model for single-label classifications. In terms of training time per epoch, we report the average time at each level of the models. The numbers we report here are computed on a single unit of GPU (NVIDIA GeForce RTX 3090, 24GB memory), on a machine with 64 units of AMD EPYC 7313 16-core Processor and 504 GB memory. We have run all the experiments on two such machines, one with 8 units of NVIDIA GeForce RTX 3090 and the other with 4 units.
For the top level, CNN takes about 6 hours, while RNN takes about 9 hours. For the second level, CNN takes 1 hour 20 minutes on average, while RNN takes 3 hour 47 minutes. For the third level, CNN and RNN take 12 and 22 minutes on average, respectively.

%For the case study of JEL, ACM and MSC we refer to the readers to Appendix \ref{app:kw-vocab}. 

The performance report suggests that CNN and RNN perform well in most cases. Taking into account the efficiency of training, CNN seems to be a clear winner to serve as a suitable ``default'' model for the entire hierarchy. For those models that have extremely poor accuracy, which we pay attention to in our improved approaches using keywords as a vocabulary list (c.f.~\cite{piriMT}) or Transformers we evaluate in the subsequent section.

\paragraph{Transformers Results.}
In the above reported results using CNN/RNN, we have 49 remaining poorly performing models that have a precision of less than 70\% even in the best performing architectures among CNN and RNN. As in previous sections, we provide a result summary here and refer readers to the full result table in our code repository. \footnote{The results of single-label experiments using Transformers are accessbile under \url{https://gitlab.ethz.ch/raox/science-clf/-/blob/main/result_tables/transformer_single_label_clean_modelnotempty_export_v2.csv}.} %Appendix \ref{app:result-full-tr-single}.

\begin{itemize}
    \item Transformers seem to improve performance in terms of precision (26 out of 49), accuracy (30 out of 49), and recall (30 out of 49). Improvement varies from as little as 0.08\% to as much as 33.27\% (from 25.19\% to 58.46\%).
    \item Due to training efficiency, we use only the batch size of 16 and run for one epoch.
    \item The average time per epoch is shorter than with RNN (sometimes only half), despite the large difference in batch size (RNN: 512 or 1,024 vs.~Transformers: 16). 
\end{itemize}

%For the case study of JEL, ACM and MSC we refer the reader to Appendix~\ref{app:transformer}. 

%%%%%
\paragraph{Summary of Single-Label Classifications.}
Although CNN/RNN models have achieved good performance, they suffer from either computational efficiency or performance, such as sequential processing in RNN. It is suggested that we can use CNN as the base model of the system, but we are in need of a superior classifier for models that are still performing poorly. Transformer models are proposed to solve the inefficiency in sequential processing, by processing the text as a whole rather than word by word sequentially. This allows for parallelization and makes processing computationally more efficient without sequential processing \citep{NIPS2017_3f5ee243}.

% R&R: break down into various parts and copy to other parts. 
\subsection{Multi-Label Experiments}
\label{sec:multi-label-eval}

After removing single-label models, we have 1,474 models to train using each of the CNN and RNN neural network architectures. 

\paragraph{CNN and RNN Results.}
As in previous sections, we provide a summary of the results here and refer the reader to the complete result table in the full result table in our code base.\footnote{The results of multi-label experiments are accessible under \url{https://gitlab.ethz.ch/raox/science-clf/-/blob/main/result_tables/multi_label_clean_modelnotempty_export_v2_fixed.csv}.} 
%in Appendix~\ref{app:result-full-multi}. 
The result summary is as follows:
% added the top level in
\begin{itemize}
    \item  Despite the conservative accuracy determination choice, the models perform acceptably: 1,123 out of 1,474 models achieve an accuracy of at least 90\% on their best models. In these models, precision and recall are equally good: 1,208 and 1,131 of 1,474 models have achieved $\gt$ 90\%, respectively.
    \item CNN continues to lead in performance in all performance metrics: It performs best in 56.7\% of the models, followed by RNN in 43.3\% of the models according to the precision score. 
    \item The average precision scores for CNN and RNN are 95.69\% and 93.18\%, respectively. This shows a small degradation from the single-label case, as expected.

\end{itemize}
The computational time for the top level is 105 hours for CNN and 1,688 hours for RNN, with a batch size of 64 due to RAM constraints. On the second level, CNN per epoch on average takes 1 hour, while RNN requires 3 hours and 14 minutes. On the third level, CNN and RNN take on average 48 minutes and 1 hour 20 minutes, respectively. Therefore, we draw conclusions similar to those of Section~\ref{sec:3lclf}: CNN is a good choice of default model in our hierarchical classification system on all levels, be it single-label or multi-label. 

%For the case study of JEL and ACM we refer the readers to the Appendix~\ref{app:multilabel}. 

\subsubsection{Transformers for Multi-Label Classification} \label{sec:transformer-multi}
Similar to the setting described in Section~\ref{sec:single-label-eval}, we also perform experiments in the Transformers on multi-label classification, whose precision is less than 70\%. There are 51 such models to train.\footnote{The results of multi-label experiments using Transformers are accessible under \url{https://gitlab.ethz.ch/raox/science-clf/-/blob/main/result_tables/transformer_multi_label_clean_modelnotempty_export_v2.csv}.} 
Here are the observations that we make. Transformers seem to improve performance in terms of precision (41 out of 51), accuracy (39 out of 51), and recall (21 out of 51); the improvement varies from as little as 0.05\% to as much as 51.9\% (from 21.01\% to 72.91\%). The training setup is identical to that in Section~\ref{sec:single-label-eval} and the training time per epoch on average also follows the pattern found in the single-label setting.

\section{Interfield Citation Scores Within and Across Disciplines} \label{sec:interdisciplinarity}

\begin{figure*}
  \centering
  \begin{tabular}{c}
    \includegraphics[width=0.93\linewidth]{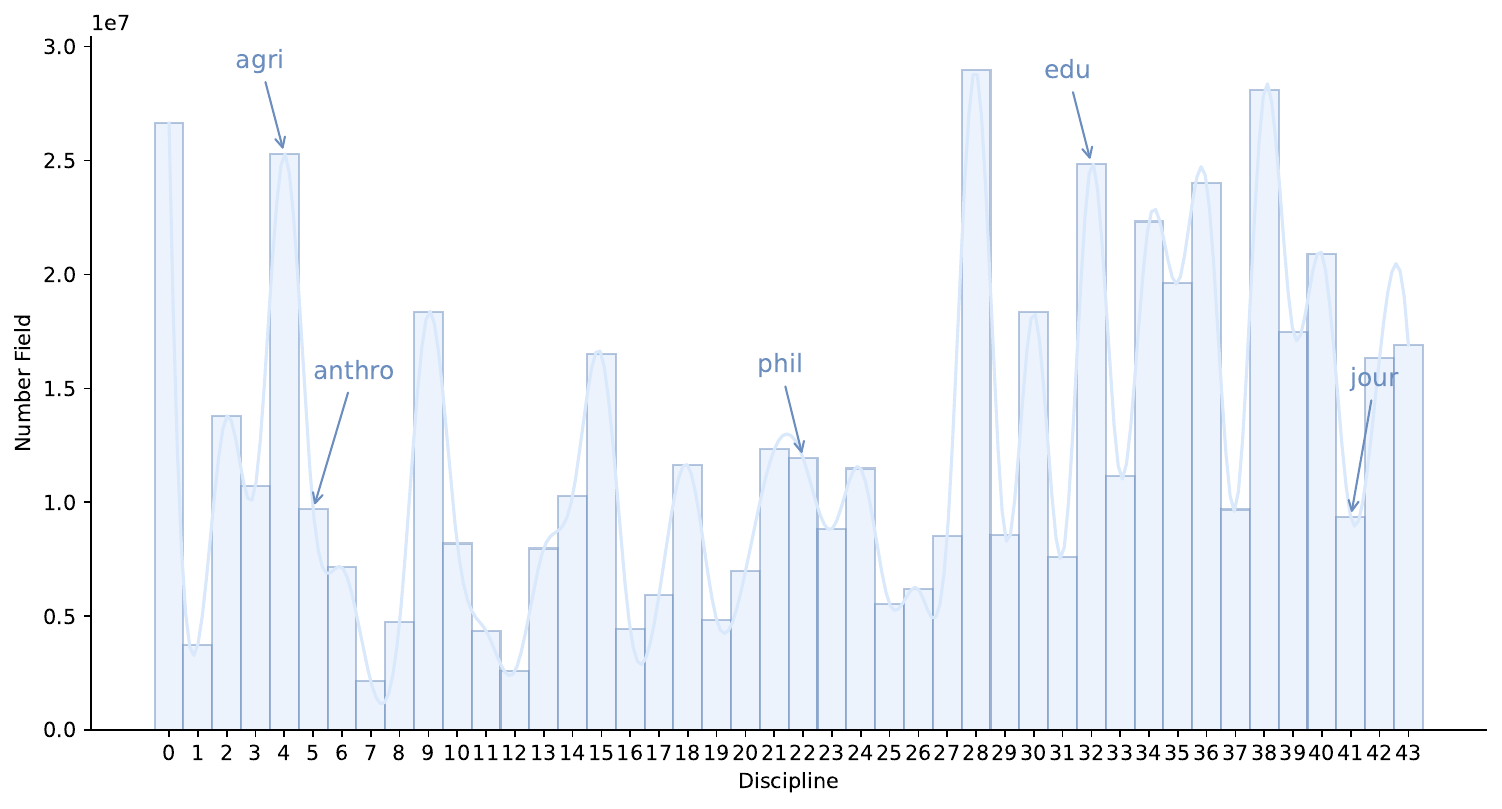} \\
    \textbf{\footnotesize (a) Number of total papers.} \\  

    \includegraphics[width=0.93\linewidth]{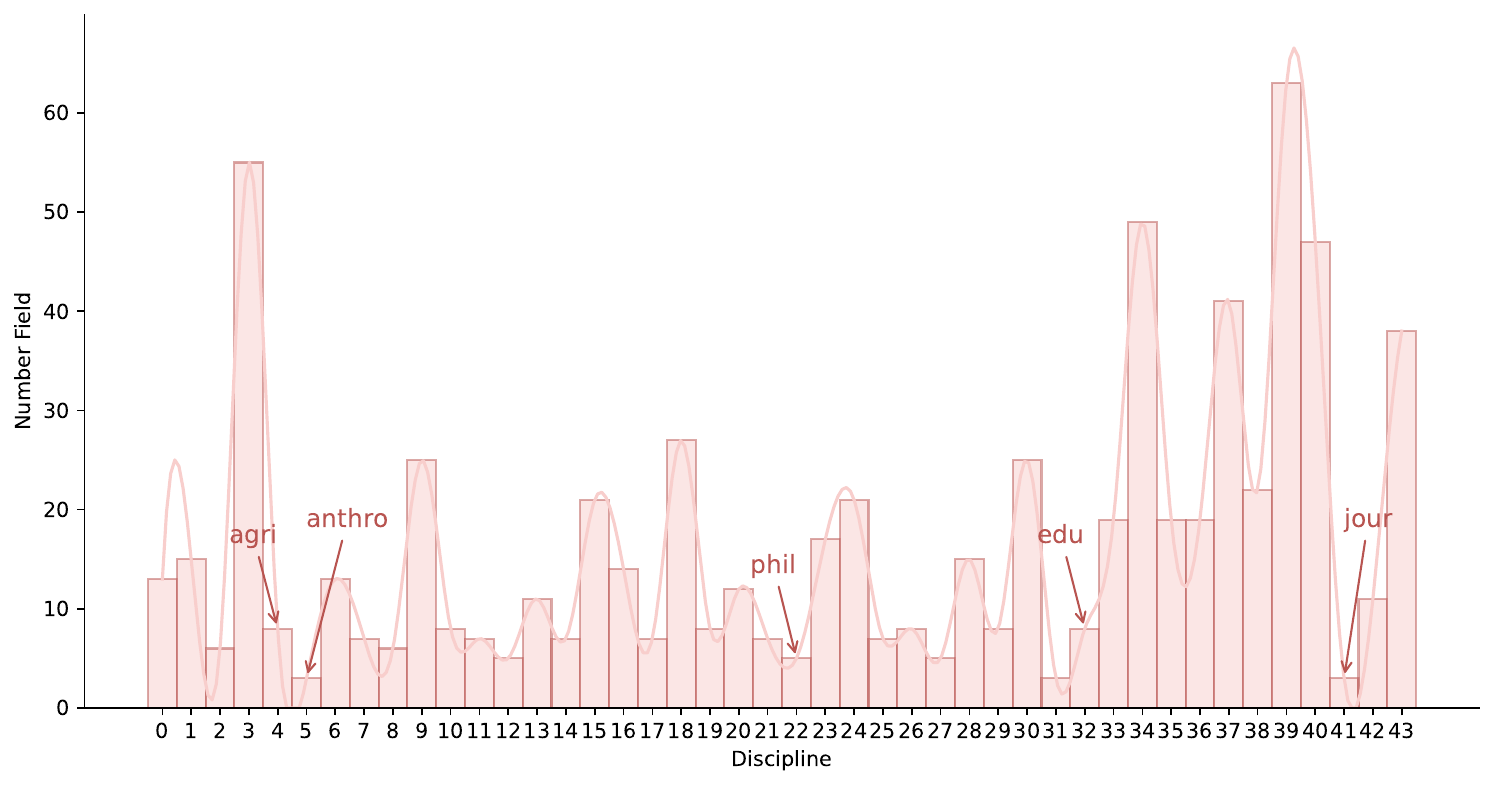} \\
      \textbf{\footnotesize (b) Number of fields.} \\
      
  \includegraphics[width=0.93\linewidth]{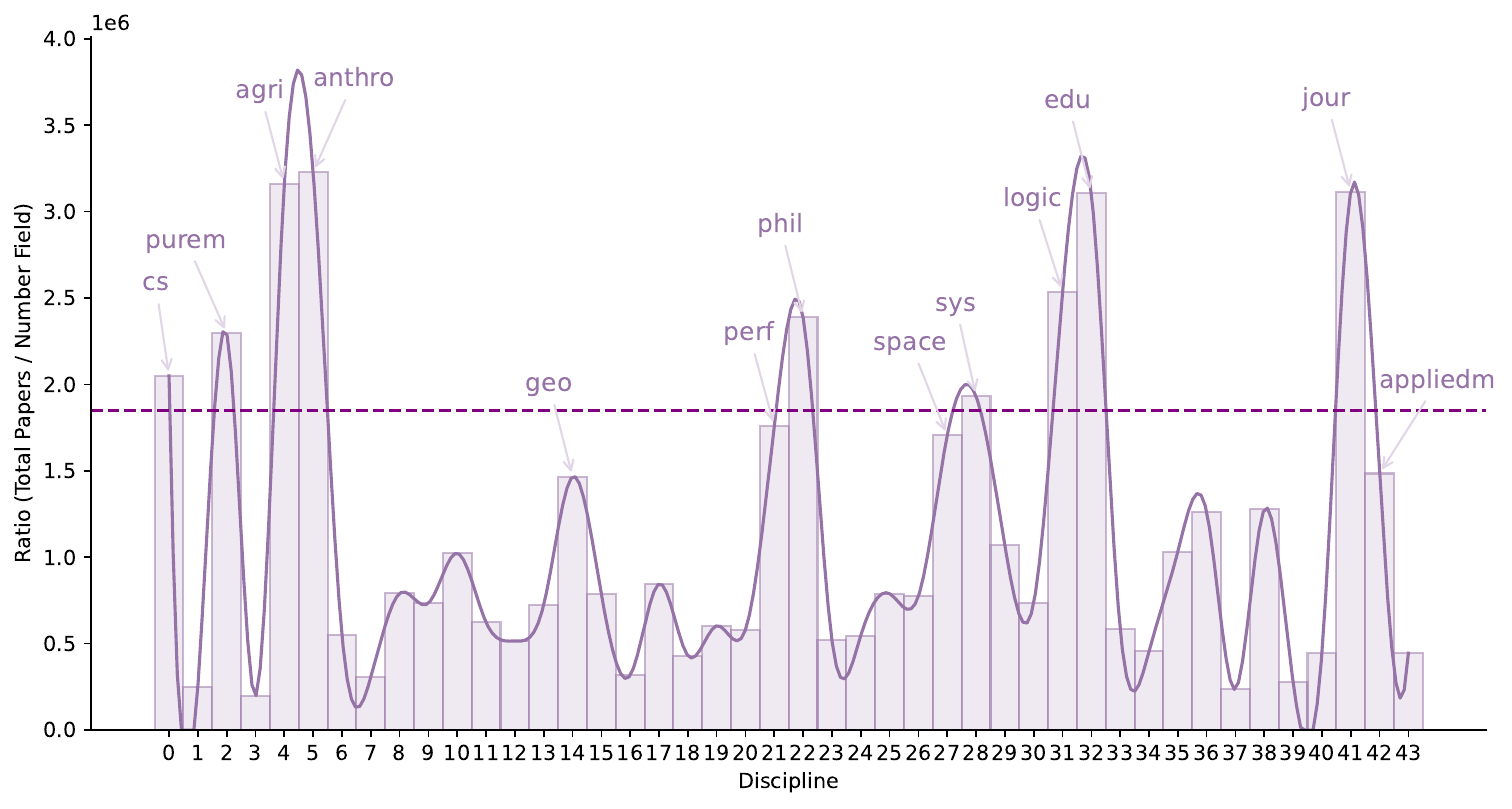} \\
  \textbf{\footnotesize (c) Ratio of total papers to number of fields.}
  
  \end{tabular}
  \caption{Distributions of total papers, number of fields and their ratio in 44 disciplines.}
  \label{fig:all_dist}
\end{figure*}

\begin{figure*}
\includegraphics[width=1\linewidth]{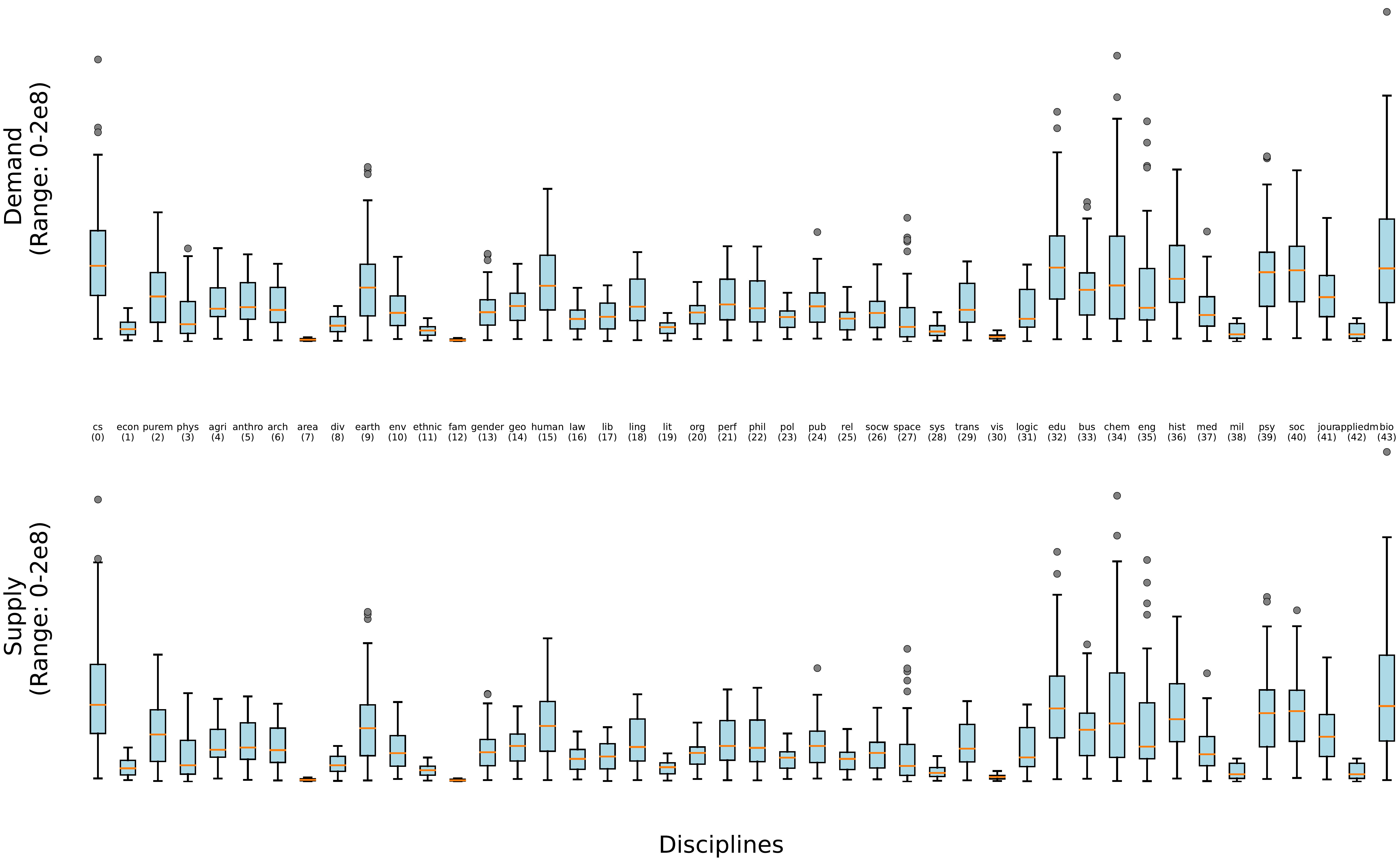}
    \caption{Box plots for demand and supply of 44 disciplines.}
  \label{fig:all_dist_box}
\end{figure*}

% - general approach
% - 44 discipline study
% - 718 field study
% - case study: 4 discipline interfield

Interdisciplinarity is in the limelight of research funding bodies and is believed to be the key to breakthrough innovations in many fields of academic interest (see  \citet{ioannidis2005, leydesdorff2009global, van2015interdisciplinary}). Measuring it requires a clear delineation of the boundaries of disciplines as well as of intra-disciplinary fields on the one hand and metrics that indicate intra- versus inter-disciplinary activity. Since disciplinary boundaries are easier to determine than those across fields within disciplines, earlier work has mainly focused on interdisciplinary rather than interfield interdependence of research. The usual interdependence metrics are citation input (demand) and output (supply) between disciplines (see \citet{ioannidis2005, leydesdorff2009global, van2015interdisciplinary}).

The framework proposed in this paper can help inform and add to this debate in the following way. First, it can identify the fields within disciplines that are particularly responsible for the absorption as well as the supply of ideas among field-to-field cells. To the extent that two fields in a pair belong to two different disciplines, this allows identification of interdisciplinary citation input and output scores. Second, the established taxonomy in this paper can determine to what extent such interdisciplinarity scores are concentrated versus dispersed between the supplying (outputting) and receiving (inputting) disciplines. Third, the same taxonomy can even help, by anchoring the authors of academic output in (potentially multiple) disciplines and fields, understanding to which extent such flows of ideas happen and interdependence is created by and from authors with a cross-disciplinary footing in their own academic work (within-author interdisciplinarity) or not (between-author interdisciplinarity). Fourth, by assigning researchers to disciplines and fields, the approach can help identify to what extent interdisciplinary impact is more or less likely in the case of interdisciplinary collaborations in academic work or not. The latter two aspects might be important for the selection of instruments geared toward promoting interdisciplinary research and its impact.

In this section, we will focus on citation inputs and outputs as metrics of influence within and across fields, as well as within and across disciplines. We will do so from a macro-perspective of disciplines (Section \ref{sec:across-discipline-44}), subsequently consider fields within disciplines (Section \ref{sec:within-discipline-718}), and finally we will offer a discussion of the latter two subsections in light of score measurement in Section \ref{sec:across-discipline-intersciplinarity}.

\subsection{General Approach} \label{sec:across-discipline-approach}
At the macro level, we will consider interdisciplinary citation matrices across all 44 considered disciplines on the one hand and all 718 fields in the disciplines on the other hand. The underlying citation linkages for each pair of outputs are provided in MAG.

Each research output (e.g., article) in MAG is classified to belong to at least one discipline of academic work and at least one field (the highest hierarchical level below the one of disciplines). Altogether, we distinguish between 44 disciplines which cover 718 fields of academic research. Hence, the average discipline covers 16 fields. Figure \ref{fig:all_dist} shows the distributions of the total academic articles (a), field numbers (b), and their ratio (c) in all 44 disciplines in our multi-label training set, which dates from 1800 to 2018.

The ratio represents the average number of articles per field within disciplines. We see that the distributions of various attributes differ largely from each other and result in the highest average numbers of articles per field (c) being ``Computer science", ``Pure mathematics", ``Agriculture", ``Anthropology",  ``Philosophy", ``System science", ``Logic", ``Education", and ``Journalism". This is due to the fact that a discipline generates many publications such as ``Computer science" or ``Agriculture", and/or it has a small number of fields such as ``Anthropology" or ``Journalism".  

Figure \ref{fig:all_dist_box} provides the whisker plots of the distribution of citations outward (supplied; output) and inward (demanded; input) in the fields for each discipline. We observe that in most of the discplines the citation supply and demand are almost equal, but their range varies largely across different disciplines. There are certain disciplines such as ``Computer science", ``Earth science", ``Gender study", ``Public policy", ``Space science", ``Education", ``Business", ``Chemistry", ``Engineering", ``Medicine", ``Psychology" and ``Biology" that have large outliers in either citation demand or supply.\footnote{For the comparison between citation demand and supply of each discipline, we refer readers to the box plots in online Appendix via \url{https://gitlab.ethz.ch/raox/science-clf/-/blob/main/result_tables/online_appendix.pdf}.}  

Let us organize the data in matrix form, so that rows are sorted first by discipline and subsequently by field within a discipline (here, we use the arbitrary numeric encodings for sorting that are fixed in the training stage, see Appendix \ref{app:discipline2coding-mapping}). This matrix is composed of 44 discipline-to-discipline blocks, where each block consists of the number of fields in the citing discipline in rows and the number of fields in the cited discipline in columns. Hence, along the diagonal, we find square blocks of intra-disciplinary citations across fields within a discipline, and off the diagonal blocks there are inter-disciplinary citations across fields between disciplines.

Let us denote the just-mentioned $718 \times 718$ matrix with citation counts by $\mathcal{I}=\mathcal{I}_{df,d^{\prime}f^{\prime}}$, where $\{d,d^{\prime}\}$ is a pair of disciplines, and $\{f,f^{\prime}\}$ are a pair of fields. In the latter statements, $d$ and $d^{\prime}$ might be identical in general, and $f$ and $f^{\prime}$ might be identical only if $d=d^{\prime}$. The matrix $\mathcal{I}$ can be though of as to be made up of discipline-to-discipline blocks that are themselves made up of field-to-field cells.

Let us use $\mathcal{I}_0$ to denote the raw matrix of citation count inputs, demand by fields in rows, and supply from fields in columns, and let $\mathcal{O}_0=\mathcal{I}^{\prime}_0$ denote its transpose, the citation count output matrix. It is customary to row normalize these matrices to focus on the distribution of counts within a row (this is also called \textit{normalization by degree}). Let us denote these normalized matrix counterparts by $\mathcal{I}$ and $\mathcal{O}$ and note that for each of them, all cell entries are nonnegative and sum up to unity in a row. Hence, the cells indicate the share of one input field (absorbing) in the output field's overall citations and the share of one output field (supplying) in the input field's overall citations, respectively. A matrix of further interest is the net output matrix $\mathcal{D}_0=\mathcal{O}_0-\mathcal{I}_0$ and its row-normalized counterpart $\mathcal{D}$, where we can use the absolute row sum normalization for the latter.

With this approach, we can consider the share of within-field citation scores (which is by definition an intra-disciplinary concept), the share of inter-field citation weights within a discipline, and the share of inter-field citation weights across disciplines. The latter is by definition an interdisciplinarity score, which can be further decomposed into specific components that accrue to individual fields. Overall, these scores provide a field-anchored description of citation gross inputs and outputs as well as net outputs.

\subsection{Interfield Citations Between 44 Disciplines} \label{sec:across-discipline-44}

First, we create a dataframe where each row is a tuple of ($Paper_1$, $Paper_2$, $Disc_1$, $Disc_2$), where $Paper_1$ and $Paper_2$ are the Paper IDs in MAG; $Disc_1$ and $Disc_2$ are the discipline labels of $Paper_1$ and $Paper_2$ in the multi-label setting. The edge value between these two paper nodes is binary, indicating a citation relationship. For instance, a tuple (2786288045, 2101095530, 43, 3) means $Paper_1$ in ``Biology" (43) with the Paper ID of 2786288045 cites $Paper_2$ in ``Physics" (3) with the Paper ID of 2101095530. We provide an extensive example in Table \ref{tab:multilabel-example} on the field level, where we have $Paper_1$ 2786288045 citing $Paper_2$ 2101095530. Note that here we provide field labels, with which we generate discipline labels. In this case, we have discipline-to-discipline mappings of (43, 3) and (43, 43), as well as field-to-field mappings (43-30, 3-18), (43-30, 43-30), and (43-30, 43-2). Each pair is an element of the Cartesian product of the label sets of $Paper_1$ and of $Paper_2$. 

In total, there are 43,718,407,275 tuples in the training set of our \textit{multi-label} settings across 44 disciplines. Then, we aggregate the tuples up into a discipline-by-discipline matrix. The discipline-to-coding mapping is listed in Appendix~\ref{app:discipline2coding-mapping}. 
% \footnote{Think of a matrix $Z_D$ which has as many rows as there are outputs, say $N_C$ and as many columns as there are disciplines, say $N_D=44$. Each row has at least one unitary entry and eventually several ones in case of multi-disciplinary associations of an output (which is enabled by the multi-label setting). Let us call the raw $N_C \times N_C $, output-to-output citation-input matrix by $C$. Then, after using $Z_D^{\prime}$ for the transpose of $Z_D$, $\mathcal{I}_0=Z_D^{\prime}CZ_D$ is a raw citation matrix of size $N_D\time N_D$ and only reflects discipline-from-discipline citation inputs. The numbers of aggregated raw citation counts are of the magnitude of $10^8$. However, we work with normalized matrices $\mathcal{I}$ and $\mathcal{O}$, as indicated above.} 

%% multilabel example

\begin{table}[]
\caption{Multi-labeled papers with field assignments (2786288045 cites 2101095530).}
\label{tab:multilabel-example}
\adjustbox{max width=1\linewidth}{
\begin{tabular}{cccc}
\toprule
\textbf{Paper ID} & \textbf{Paper title} & \textbf{Field labels} & \textbf{Field} \\
\midrule
2786288045 & \begin{tabular}[c]{@{}c@{}}Chapter 1 – Definition of Gastroesophageal Reflux \\ Disease: Past, Present,   and Future\end{tabular} & 43-30        & Bio-Pathology \\
\midrule
\multirow{3}{*}{2101095530} & \multirow{3}{*}{\begin{tabular}[c]{@{}c@{}}Esophageal Adenocarcinoma Incidence: \\ Are We Reaching the Peak?\end{tabular}} & 3-18 & PACS-Biological and medical physics \\
 & & 43-30 & Bio-Pathology \\
 & & 43-2 & Bio-Endocrinology \\
 \bottomrule
\end{tabular}
}
\end{table}

Normalization by row sum of matrix $\mathcal{I}_0$ describes the ratio of all disciplines in terms of demand from a specific discipline $d$, while normalization by row sum of matrix $\mathcal{O}_0$ (which corresponds to column normalization of $\mathcal{I}_0$) describes the ratio of supply from all disciplines to a specific discipline $d^\prime$. 

%% 44d start %%%
\begin{figure*}
  \centering
  \begin{tabular}{c}
  \textbf{\tiny (a) Input matrix $\mathcal{I}.$ } \\
  \includegraphics[width=0.93\linewidth]{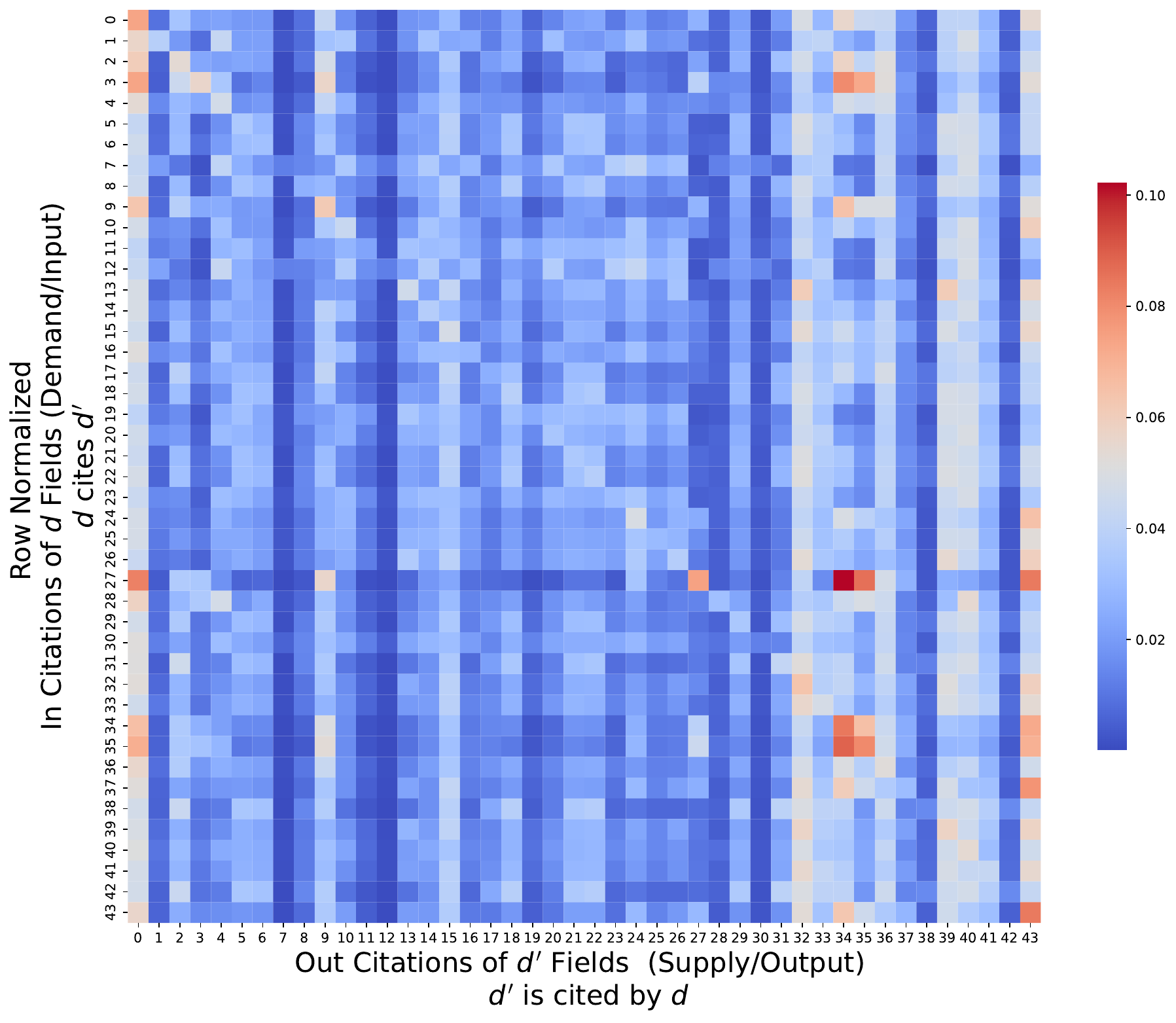} \\
  \textbf{\tiny (b) Output matrix $\mathcal{O}.$} \\
  \includegraphics[width=0.93\linewidth]{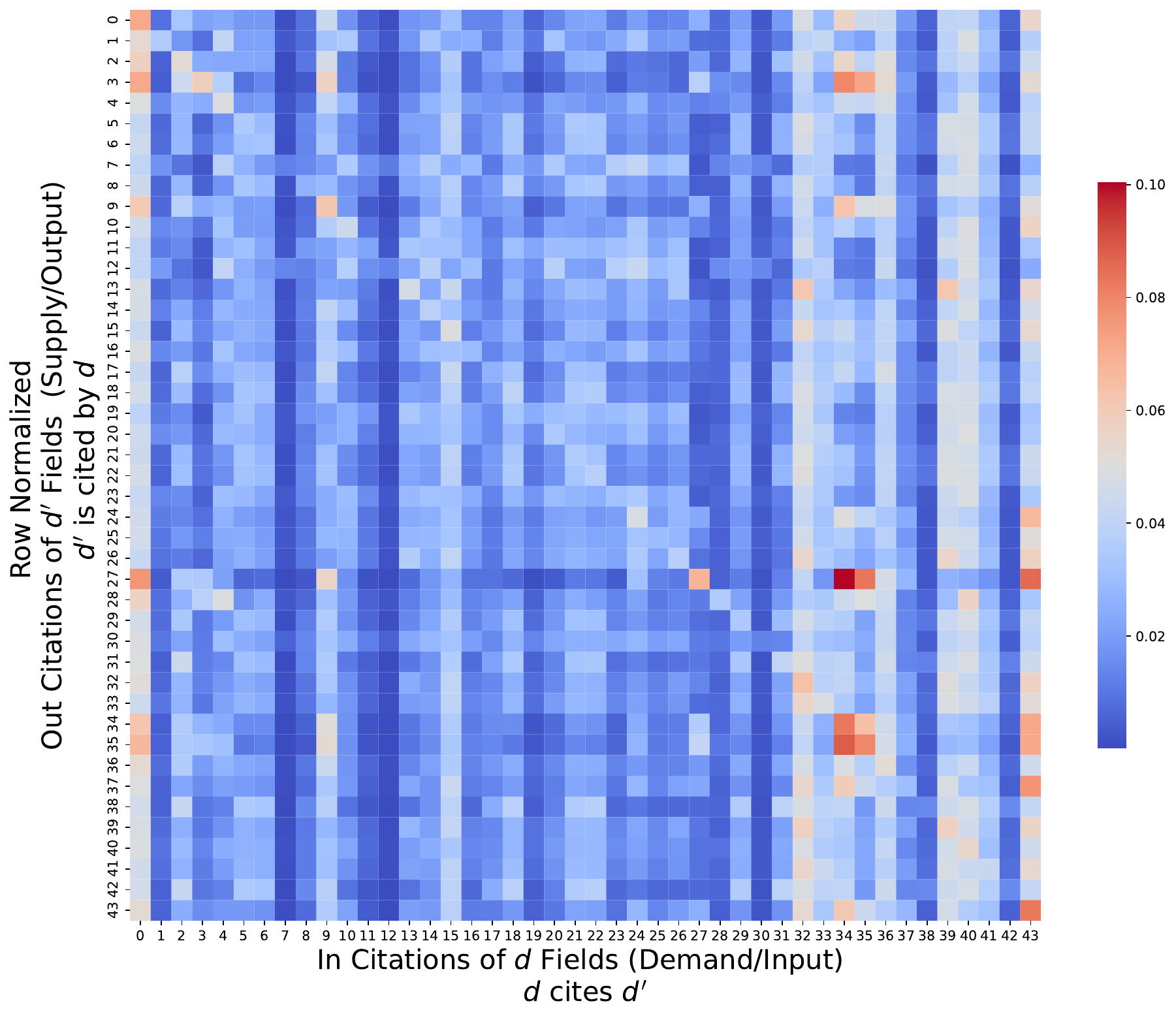} \\
  \end{tabular}
  \caption{Across-discipline interdisciplinary scores of all 44 disciplines.}
  \label{fig:out_normalized}
\end{figure*}

%%% 44d end %%%

Based on the normalized matrices $\mathcal{I}$ and $\mathcal{O}$, we generate heatmaps in panels (a) for the input matrix $\mathcal{I}$ and (b) the output matrix $\mathcal{O}$ in Figure \ref{fig:out_normalized}. The heatmaps permit identifying disciplines that face a high \textit{relative} interdisciplinary demand from one discipline (in a row across columns of $\mathcal{I}$) and ones that generate a high \textit{relative} output across disciplines from a given discipline  (in a row across columns of $\mathcal{O}$). Interestingly, the right-stochastic matrices $\mathcal{I}$ and $\mathcal{O}$ do not show dominating diagonal elements (dominance of intra-disciplinary impact) throughout the disciplines, but some disciplines are more strongly represented in interdisciplinary input dependence and output impact than others.

Disciplines that have a particularly high interdisciplinary demand (red cells in rows in Figure~\ref{fig:out_normalized} (a)) are the following:
\begin{itemize}
    \item ``Computer science" (0) highly demands (cites heavily) from ``Chemistry" (34) and ``Biology" (43).
    \item ``Physics" (3) has a high demand from ``Computer sciences" (0), ``Earth sciences" (9), ``Chemistry" (34) and ``Engineering and technology" (35).
    \item ``Earth sciences" (9) have high demands from ``Chemistry" (34).
    \item ``Space sciences" (27) highly demands from ``Computer science" (0), ``Earth sciences" (9), ``Chemistry" (34), ``Engineering and technology" (35) and ``Biology" (43).
    \item ``Chemistry" (34) demands highly from ``Engineering and technology" (35) and ``Biology" (43).
    \item ``Engineering and technology" (35) demands highly  from ``Chemistry" (34) and ``Biology" (43).
\end{itemize}

Disciplines that have a high interdisciplinary supply (red cells in rows of Figure \ref{fig:out_normalized} (b)) are the following:
\begin{itemize}
    \item ``Computer science" (0) is highly cited by ``Chemistry" (34) and ``Biology" (43).
    \item ``Physics" (3) is highly cited by ``Computer sciences" (0), ``Earth science" (9), ``Chemistry" (34), ``Engineering and technology" (35), and ``Biology" (43).
    \item ``Earth sciences" (9) is highly cited by ``Computer science" (0) and ``Chemistry" (34).
    \item ``Gender studies" (13) are highly cited by ``Education" (32), ``Psychology" (39) and ``Biology" (43).
    \item ``Space sciences" (27) is highly cited by ``Computer science" (0), ``Chemistry" (34), ``Engineering and technology" (35) and ``Biology" (43).
    \item ``Chemistry" (34) is highly cited by ``Computer science" (0), ``Earth science" (9), ``Engineering and technology" (35) and ``Biology" (43).
    \item ``Engineering and technology" (35) is highly cited by ``Computer science" (0), ``Earth sciences" (9), ``Chemistry" (34), and ``Biology" (43).
    \item ``Medicine" (37) is highly cited by ``Education" (32), ``Chemistry" (34) and ``Biology" (43).
    \item ``Psychology" (39) is highly cited by ``Education" (32) and ``Biology" (43).
    \item ``Biology" (43) is highly cited by ``Computer science" (0), ``Education" (32), ``Chemistry" (34), and ``Psychology" (39).
    \item ``Computer science" (0), ``Education" (32), ``History" (36), ``Psychology" (39), ``Sociology" (40), and ``Biology" (43) are disciplines that are cited by almost all disciplines.
\end{itemize}

In general, the above analysis suggests two interesting findings. First, in many disciplines, contributions from other disciplines outweigh those within the discipline. Second, often the relative importance of ``foreign'' disciplines in terms of demand and supply of impact is often but not always mutual. These findings are consistent with the conclusions drawn from previous research discussed in \citet{ioannidis2005, leydesdorff2009global, van2015interdisciplinary}. However, the present findings are drawn from much more comprehensive sets of fine-grained data.

\begin{figure}
    \centering
    \includegraphics[width=0.8\linewidth]{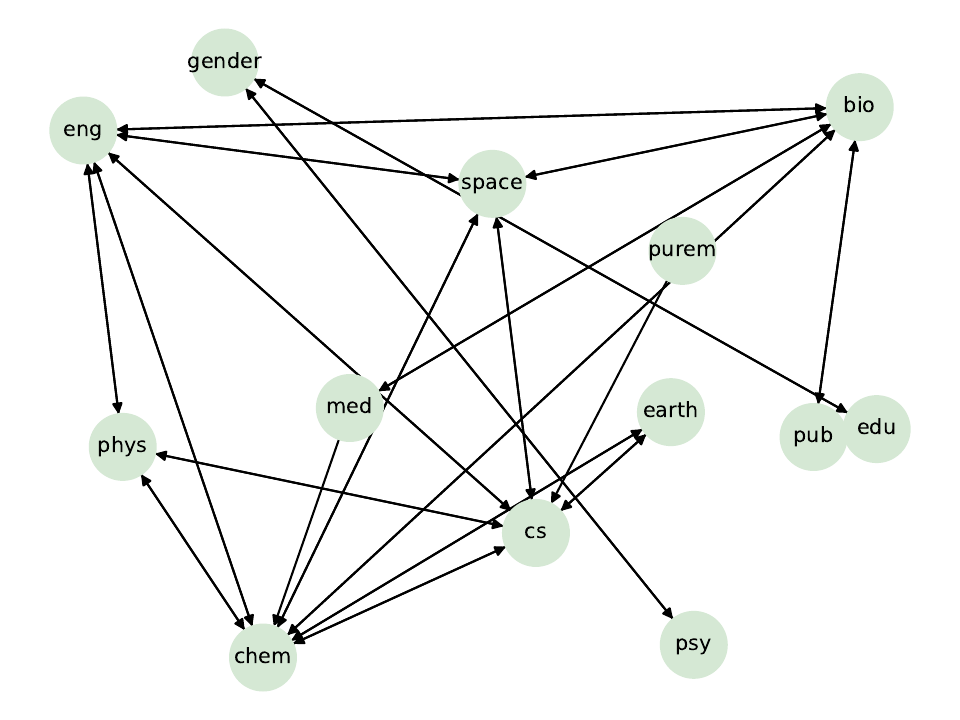}
    \caption{Diagram of high demand (out-degree) and high supply (in-degree) among selected disciplines.  \tiny Discipline labels are ``eng": Engineering and technology; ``space": Space sciences; ``pub": Public policy; ``gender": Gender and sexuality studies; ``earth": Earth sciences; ``bio": Biology; ``purem": Pure mathematics; ``cs": Computer science; ``edu": Education; ``phys": Physics; ``chem": Chemistry; ``med": Medicine; ``psy": Psychology.}
    \label{fig:out_normalized_diagram}
\end{figure}

\begin{figure*}
    \centering
    \begin{tabular}{c}
         \tiny \textbf{(a) $\mathcal{D}_0$} \\
         \includegraphics[width=0.93\linewidth]{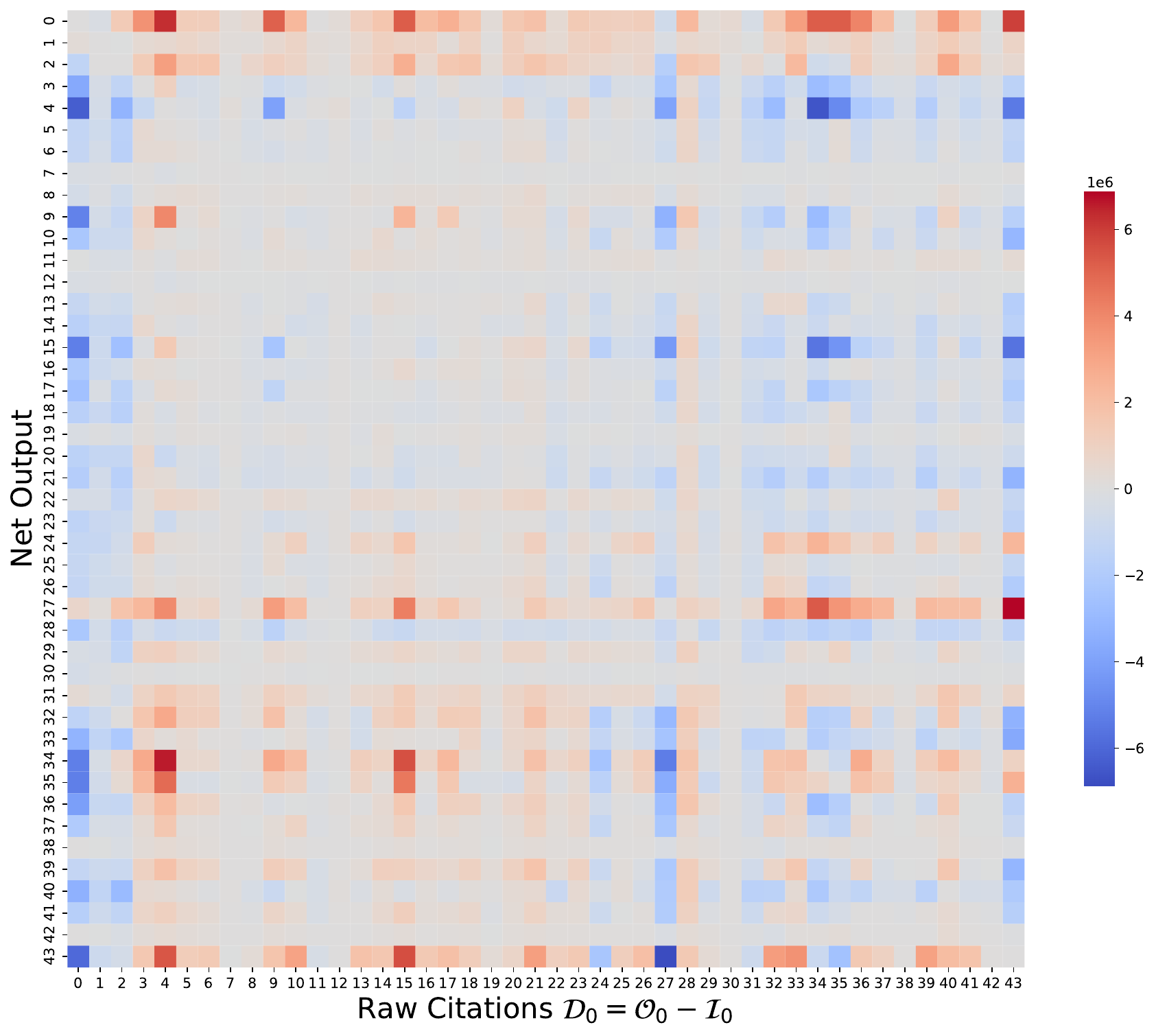}\\
         \tiny \textbf{(b) $\mathcal{D}$}\\
         \includegraphics[width=0.93\linewidth]{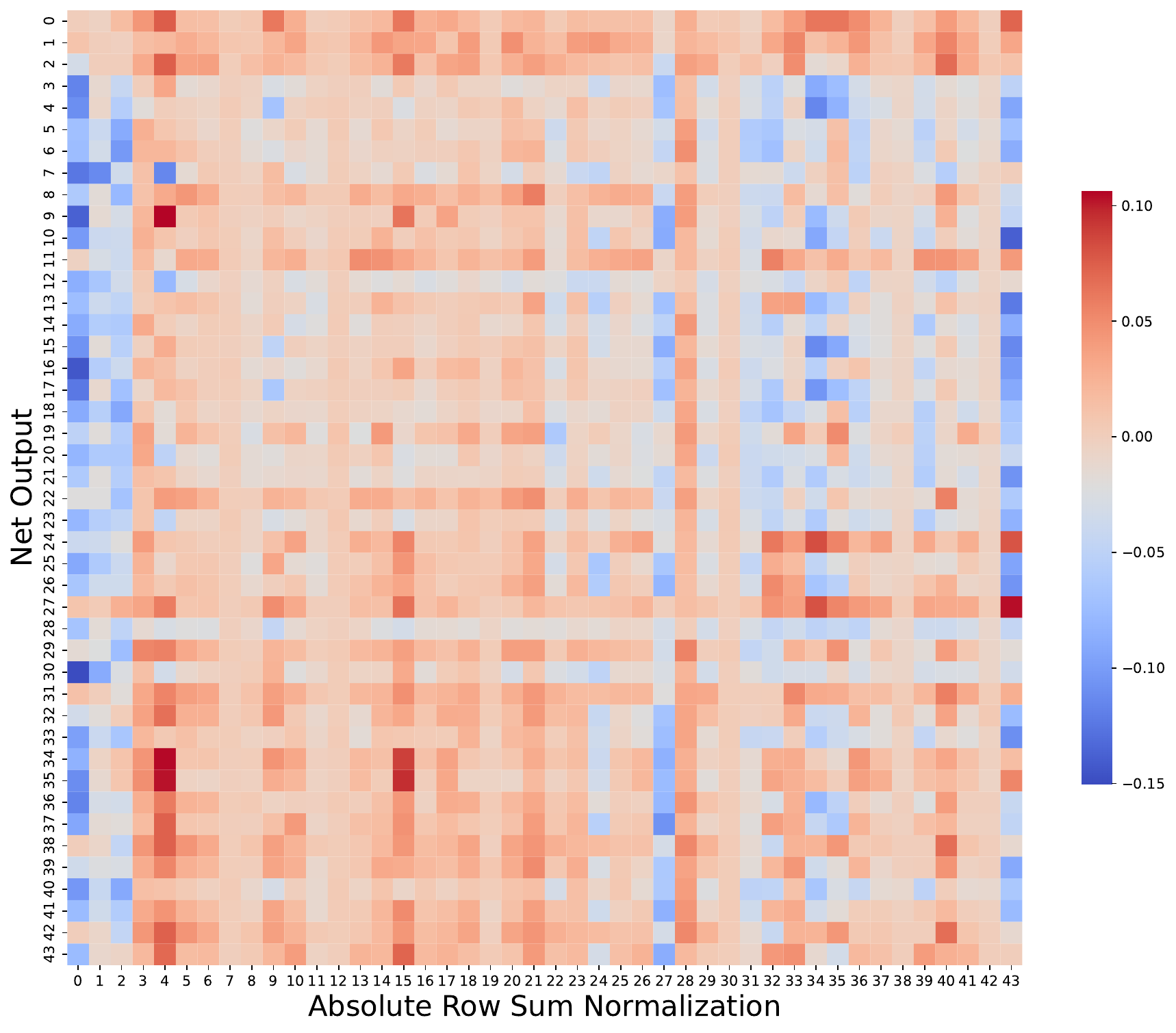}\\
    \end{tabular}
    \caption{Net output matrices (raw net citation counts and row-normalized). }
    \label{fig:D_44}
\end{figure*}

To better visualize the strong demand-supply relationships among some disciplines, we only look at red cells (with relative row-normalized citation values greater than 0.06) in Figure \ref{fig:out_normalized} and draw them in a diagram. In Figure~\ref{fig:out_normalized_diagram}, we summarize the above-mentioned finding in a bidirectional graph, where the out-degree indicates demand and the in-degree indicates supply of impact. For instance, when ``Computer science" ($d$) cites (demands from) ``Chemistry" ($d^{\prime}$), we speak of the relative out-degree of ``Computer science" ($d$), here. Conversely, when ``Computer science" ($d$) is cited by (supplies to) ``Chemistry" ($d^{\prime}$), we speak of relative in-degree of ``Computer science" ($d$).

We observe in Figure~\ref{fig:out_normalized_diagram} that ``Engineering and Technology", ``Computer science",  ``Chemistry", and ``Biology" have both a high in-degree (supply) and a high out-degree (demand). Moreover, some disciplines such as ``Medicine", ``Chemistry", and ``Biology" form a relative citation cluster with mutual interdependencies (by counting the triangular structures). It is essential to understand these connections and dependencies. For instance, they may reveal an intrinsic similarity among disciplines in terms of the research methods and subjects (reflected in mutual impact) apart from interdisciplinary influence in a more narrow sense.

We now discuss the net output matrix $\mathcal{D}_0 =\mathcal{O}_0-\mathcal{I}_0$, where $\mathcal{I}_0$ denotes the raw matrix of citation-count inputs, demand by fields in rows and supply from fields in columns and $\mathcal{O}_0=\mathcal{I}^{\prime}_0$.\footnote{With matrices and vectors, a superscript prime ($\prime$) will generally indicate a transpose.} Its row-normalized variant is $\mathcal{D}$. They are illustrated in Figure \ref{fig:D_44}. Note that $\mathcal{D}_0$ is normalized by the absolute row sum.\footnote{Note that one should not simply row-normalize $\mathcal{D}_0$, because $\mathcal{D}$ contains positive and negative cell entries.} An inspection of $\mathcal{D}_0$ suggests that certain disciplines have a positive net output, which is indicated by the red rows in Figrue \ref{fig:D_44} (a), e.g., ``Computer science" (0), ``Economics" (1), ``Pure mathematics" (2), ``Space science" (27), ``Chemistry" (34) and ``Biology" (43). The latter means that a discipline influences other disciplines more than it absorbs from them.

\subsection{Intra- And Interfield Citation Scores For 718 Fields Within 44 Disciplines} \label{sec:within-discipline-718}

%%% 718 start %%%
\begin{figure*}
  \centering
  \includegraphics[width=\linewidth]{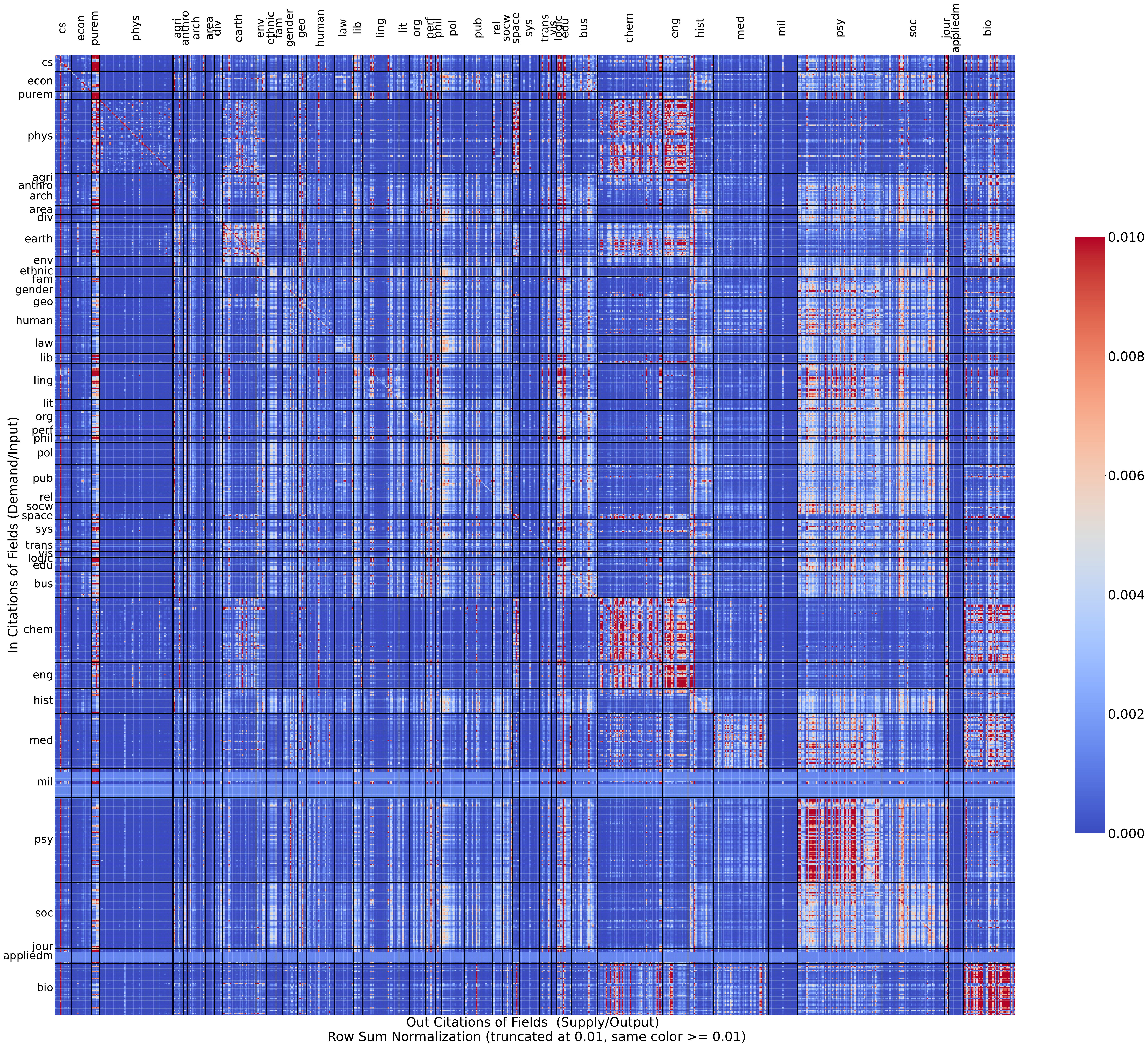} \\

  \caption{Across-and-within field interdisciplinary \textbf{row-normalized} demand of all fields in $\mathcal{I}$.}
  \label{fig:f718_row}
\end{figure*}

\begin{figure*}
  \centering

  \includegraphics[width=\linewidth]{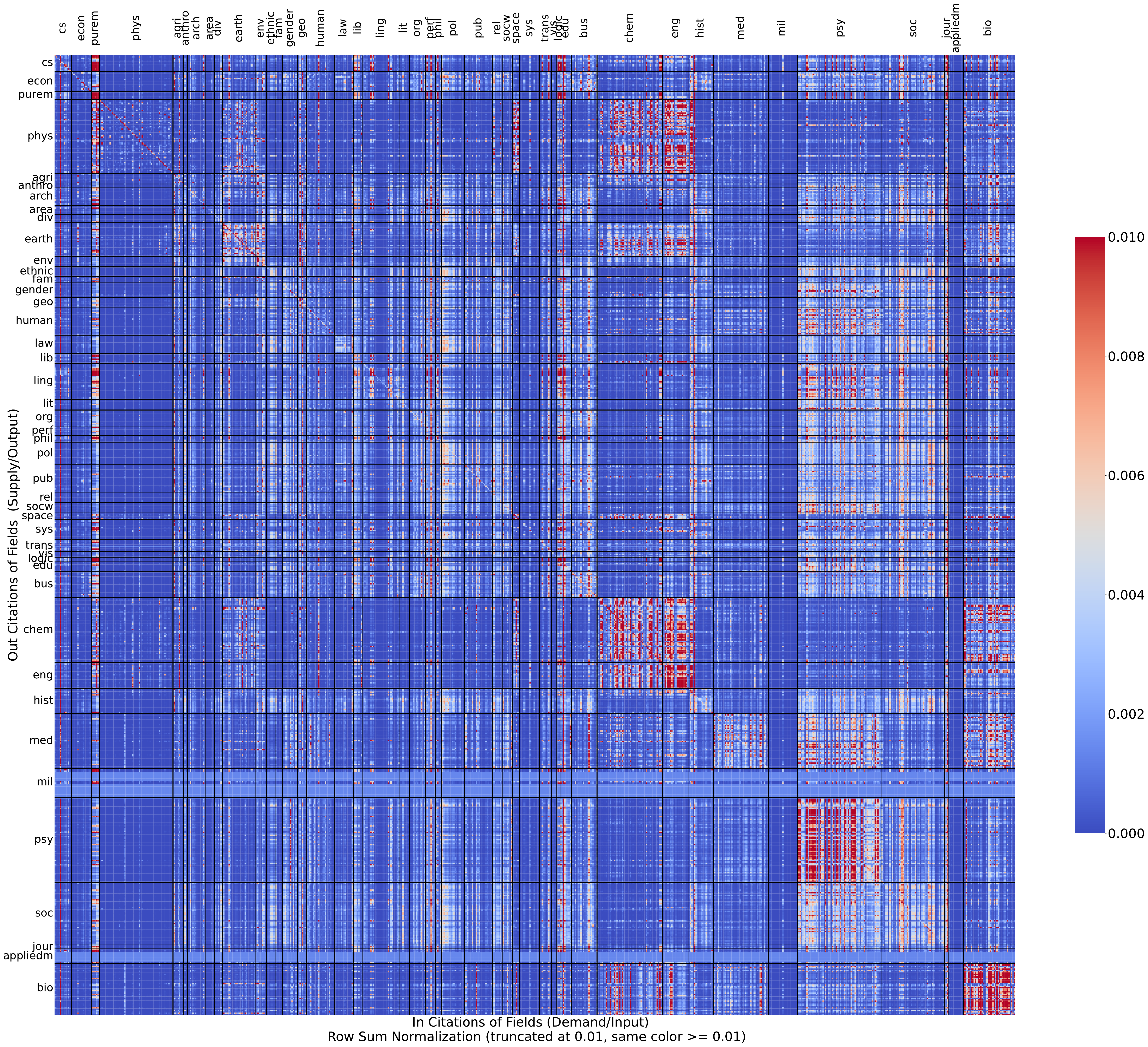} \\
  \caption{Across-and-within field interdisciplinary \textbf{row-normalized} supply of all fields in $\mathcal{O}$.}
  \label{fig:f718_col}
\end{figure*}

\begin{figure*}
    \centering
        \includegraphics[width=1\linewidth]{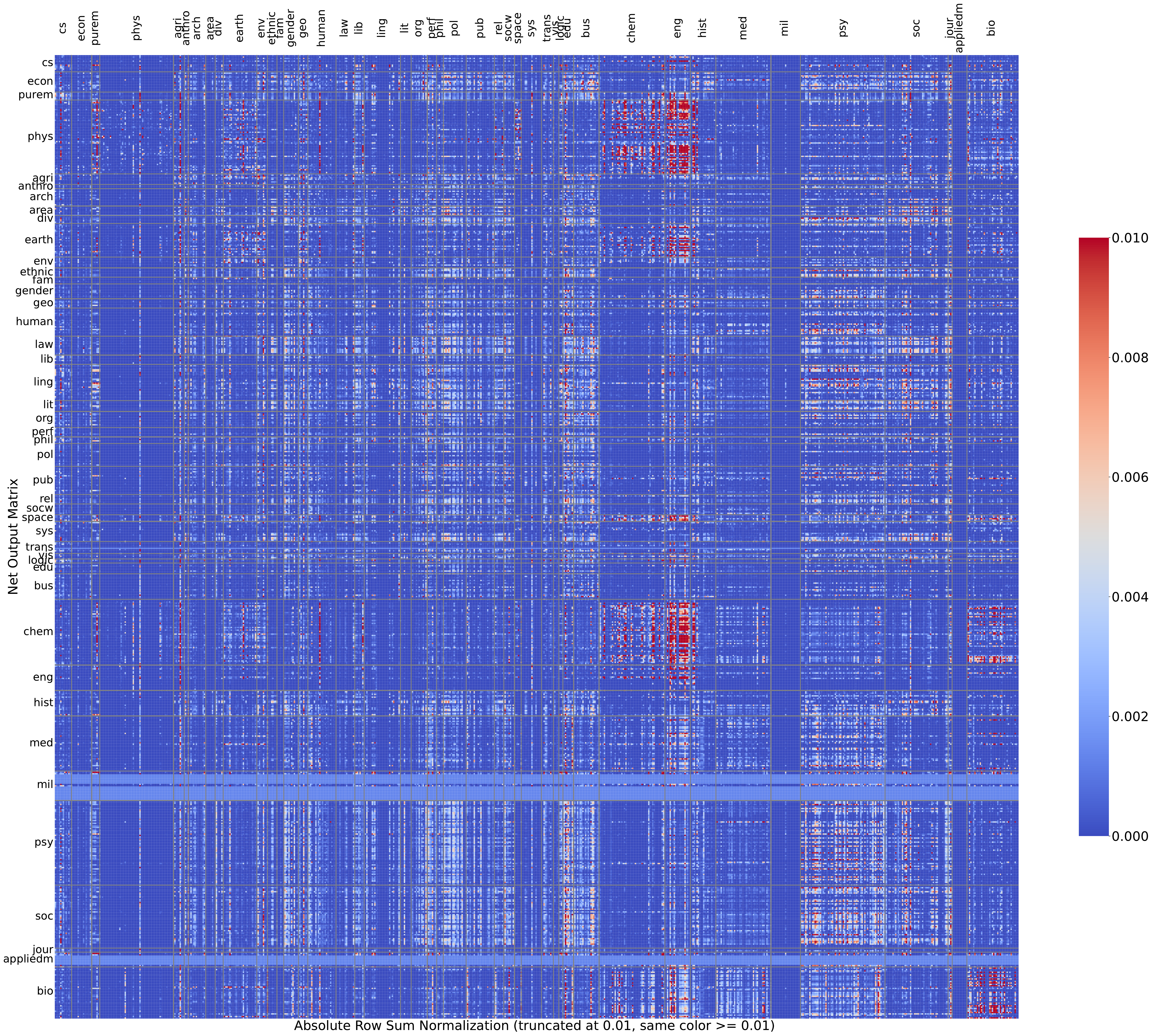}\\

    \caption{Truncated absolute row-normalized net output matrix $\mathcal{D}$ across 718 fields.}
  \label{fig:f718_D_norm_truncated}
\end{figure*}

We next move from pairs $\{d, d^\prime\}$ of disciplines to pairs $\{f, f^\prime\}$ of fields.%\footnote{Analogous to disciplines, use a matrix $Z_F$ which has $N_C$ rows for outputs and as many columns as there are fields, say $N_D=718$. Again, each row has at least one unitary entry. Then, $\mathcal{I}_0=Z_F^{\prime}CZ_F$ is one raw citation matrix of size $N_D\time N_D$ and only reflects discipline-from-discipline citation inputs. The field-to-coding mapping is listed in Appendix~\ref{app:discipline2coding-mapping}.} 
Figure \ref{fig:f718_row} illustrates the distribution of row-normalized inter-field impacts among 718 fields, focusing on relative citation input/demand. For better visibility, we do not only row normalize the matrices $\mathcal{I}_0$ and $\mathcal{O}_0$ but additionally use a truncation threshold value of 0.01 (setting all cell values larger than the threshold value to 0.01). Moreover, we indicate the disciplinary boundaries so as to clearly spot intra- and inter-disciplinary demand across fields. We can summarize the respective results as follows:

\begin{itemize}
    \item Within each discipline,  fields in ``Physics", ``Earth science", ``Logic", ``Chemistry", and ``Biology" have a strong intra-field citation demand.
    \item Some disciplines have extremely high intra-disciplinary inter-field impacts compared to inter-disciplinary inter-field impacts. Examples are ``Psychology" and ``Biology". 
    \item There is only one discipline, ``Physics", where the strongest inter-field citation demand is not from within its own discipline, but from outside (namely ``Space science", ``Chemistry" and ``Engineering").
    \item There are some disciplines such as ``Psychology" and ``Sociology" that are highly demanded by clusters of disciplines (those belonging to the humanities and social sciences).
\end{itemize}

Figure \ref{fig:f718_col} focuses on the (truncated) relative citation output/suply. For reasons of better visibility, here we also choose a threshold value of 0.01. The corresponding findings can be summarized as follows:

\begin{itemize}
    \item Intra-field supply in most fields is not higher compared to the supply from other fields within the same discipline.
    \item All fields in some disciplines supply to many fields in some related discipline. E.g,. fields in ``Physics" supply highly to those in ``Chemistry", ``Space science", and ``Engineering".
    \item ``Psychology" and ``Biology" display a high inter-field disciplinarity in supply within their disciplinary boundaries.
\end{itemize}

%% about D and its row-normalized matrix

Following the procedure for computing the net output matrix $\mathcal{D}_0$ and its absolute row normalized variant $\mathcal{D}$  in Section \ref{sec:across-discipline-44}, we generate a normalized net output matrix of fields. To better visualize the matrix $\mathcal{D}$, we again truncate it, using a threshold value of 0.01 for the cell entries of $\mathcal{D}$ in Figure \ref{fig:f718_D_norm_truncated}.  

From this we observe that fields in disciplines such as ``Physics", ``Earth science", ``Chemistry", and ``Biology" form a cluster that generates net citation output to fields from the same cluster but in other disciplines. 

We see that the fields in ``Physics" and ``Chemistry" have a strong impact on the fields in ``Chemistry" and ``Engineering" (red cells). We see that there are no largely excessive net relative inputs (negative net relative output) relative to the chosen threshold value. However, if we look at the net output at the discipline level (Figure \ref{fig:D_44}), the variation is much greater.

\subsection{Interfieldness Within and Across Disciplines (Interdisciplinarity)} \label{sec:across-discipline-intersciplinarity}
We care about various levels of the impact of a research field or a discipline on others. In what follows, we will focus on ``interfieldness'' -- the citation connectivity between pairs of fields -- within a discipline (Section \ref{sec:within-discipline}) and across disciplines, that is, interdisciplinarity (Section \ref{sec:across-discipline}), always keeping the focus on a granularity at the level of fields of research.

\subsubsection{Within-Discipline Interfieldness} \label{sec:within-discipline}

\begin{figure*}[]
    \centering
    \includegraphics[]{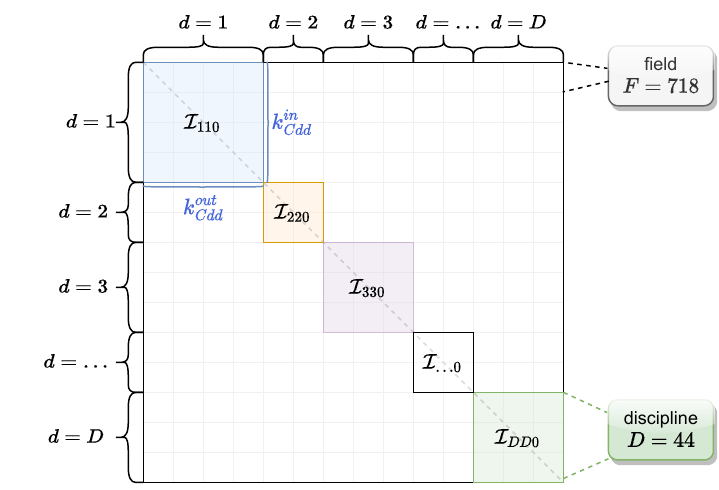}
    \caption{Illustration of interfieldness and units of analysis.}
    \label{fig:interfieldness-vis}
\end{figure*}

% Two figures like Fig 13. in (CS / Econ) showing outside demand vs. outside supply, they need to be sufficiently different from each other 
\begin{figure*}[]
    \begin{tabular}{c}
        \includegraphics[width=1\linewidth]{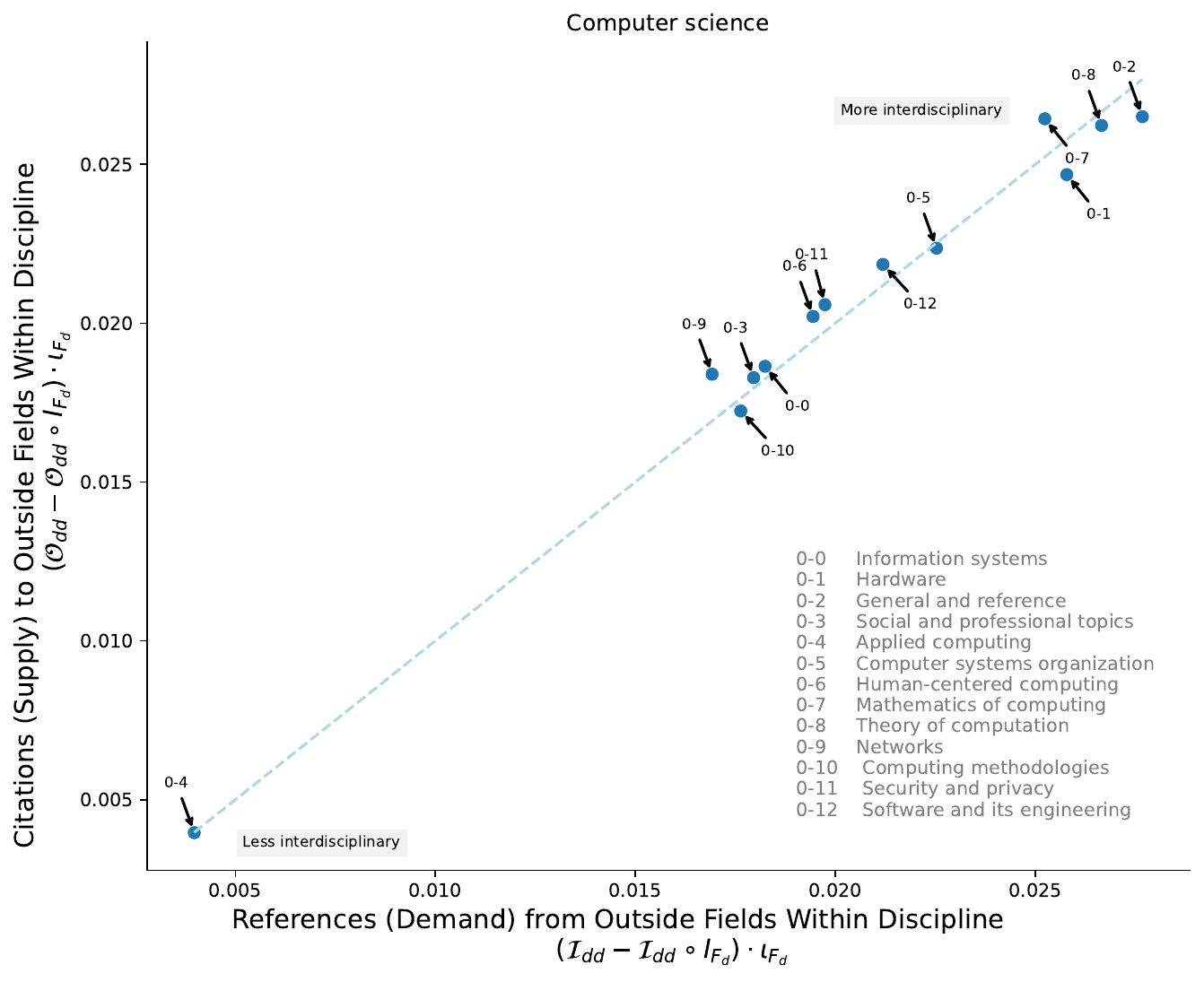} \\
        \includegraphics[width=1\textwidth]{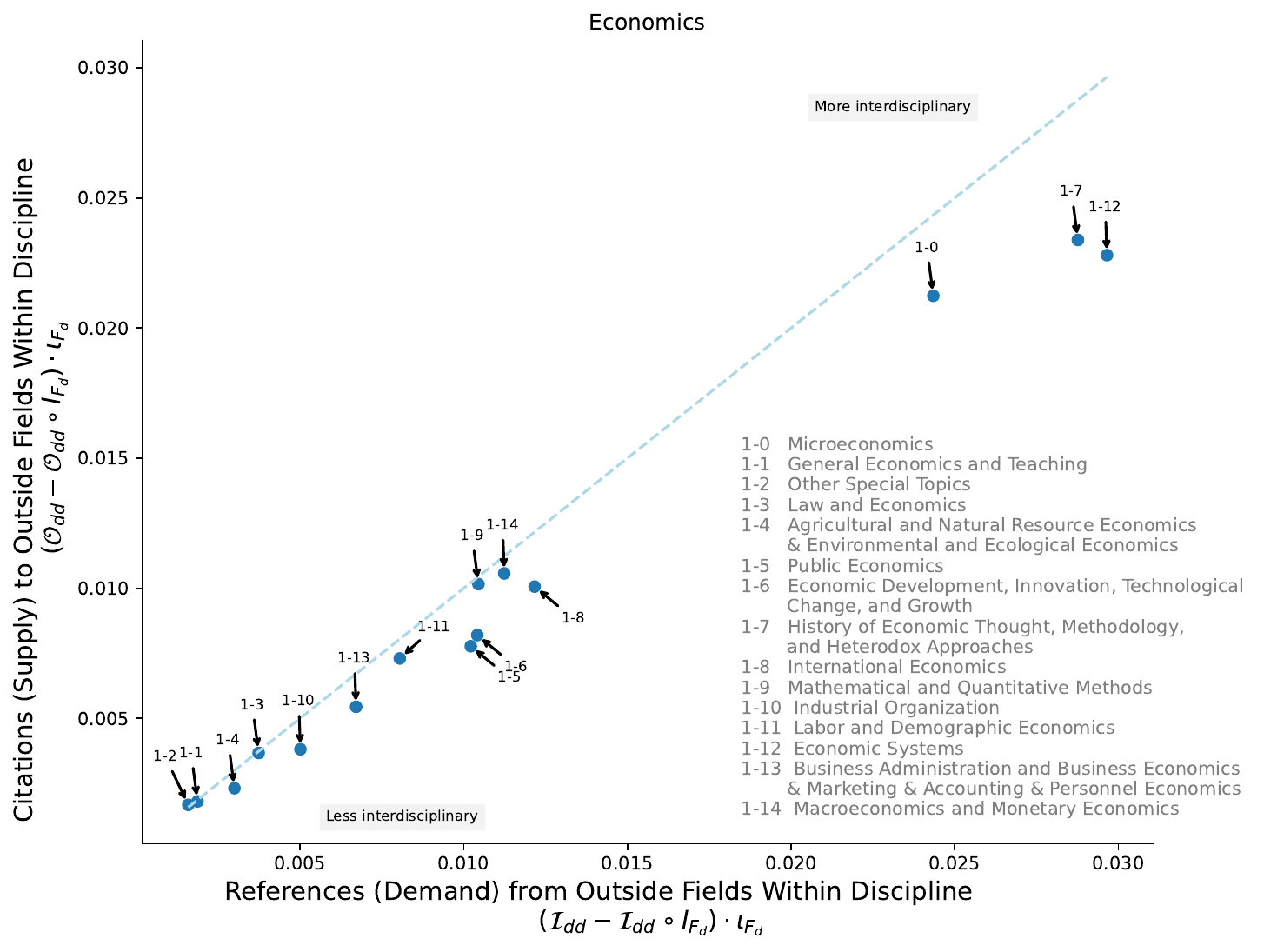} \\
    \end{tabular}
    \caption{Interfieldness by demand from and supply to other within-discipline fields in two exemplary disciplines (``Computer science" and ``Economics").}
    \label{fig:scatterplot_cs_econ}
\end{figure*}

Here, we will focus on one concept of measurement of interfieldness: normalized cross-field citation counts within a discipline. Think of an $N_{d}\times N_{F_d}$ assignment matrix $Z_{F_d}$ that has as many rows as there are publications in discipline $d$ and as many columns as there are fields in $d$. Let $C_{dd}$ be the binary-entry, $N_d\times N_d$ citation-input matrix of field $d$. Then, we obtain an unnormalized interfieldness matrix of size $F_d\times F_d$:
\begin{eqnarray}
\mathcal{I}_{dd0}&=&Z^{\prime}_{F_d}C_{dd}Z_{F_d},
\end{eqnarray}
where $\mathcal{I}_{dd0}$ is a submatrix of the earlier citation-input matrix $\mathcal{I}_0$ for all pairs $\{f,f^{\prime}\}$ of fields in discipline $d$. Note that, as before, the raw output matrix is defined as $\mathcal{O}_{dd0}=\mathcal{I}^{\prime}_{dd0}$. Discipline block matrices $\mathcal{I}_{dd0}$ along the diagonals are visualized in Figure \ref{fig:interfieldness-vis}. Again, we row normalize $\mathcal{I}_{dd0}$ to get $\mathcal{I}_{dd}$ and row normalize $\mathcal{I}^{\prime}_{dd0}$ to get $\mathcal{O}_{dd}$.

For any generic matrix $V$ with elements $v_{ij}$, let $tr(V)=\sum_i v_{ii}$ denote the trace of $V$, defined as the sum of diagonal elements. With an identity matrix $I_{N_d}$ that has the same number of rows and columns as there are publications in the discipline $d$, $N_d$, we have $tr(I_{N_d})=N_d$. Note that by design, $tr(C_{dd})=0$, because all citations are \textit{between} research publications (there are no self-citations at the publication level). Let $\iota_{N_d}$ and $\iota_{F_d}$ denote column vectors of as many rows as there are publications and fields in the discipline $d$, respectively. It will be useful to utilize $k$ for $F_d\times 1$ vectors of (some) degree in rows of matrices and $\kappa$ for degree-based scalars.

Armed with those definitions, we can state the following properties for each field:
\begin{itemize}
\item the vector of the total number of in-citations per field within discipline $d$ is $k^{in}_{Cdd}=\mathcal{I}_{dd0}\iota_{F_d}$;
\item the vector of the total number of out-citations per field within discipline $d$ is $k^{out}_{Cdd}=\mathcal{O}_{dd0}\iota_{F_d}$;
\item the total number of citations within discipline $d$ is $\kappa^{total}_{Cdd}=\iota^{\prime}_{F_d}\mathcal{I}_{dd0}\iota_{F_d}$, where $\kappa^{total}_{Cdd}=\iota^{\prime}_{F_d}k^{in}_{Cdd}=\iota^{\prime}_{F_d}k^{out}_{Cdd}$;
\item the total number of intrafield citations in discipline $d$ is $\kappa^{intra}_{Cdd} $ $=tr(\mathcal{I}_{dd0})=tr(\mathcal{O}_{dd0})$;
\item the total number of interfield citations in discipline $d$ is $\kappa^{inter}_{Cdd} $ $=\kappa^{total}_{Cdd}-\kappa^{intra}_{Cdd}$.
\end{itemize}

We first focus on the interfieldness within each discipline, i.e., the intra-disciplinary but interfield citation scores. By subtracting the diagonal elements in the normalized matrices $\mathcal{I}_{dd}$ and $\mathcal{O}_{dd}$ (i.e., the ratio of intrafield citations $\kappa^{intra}_{Cdd}$ in $k^{in}_{Cdd}$ and $k^{out}_{Cdd}$), we plot the demand and supply of fields from and to other fields within a discipline. An inspection of those plots attests to varying patterns of interfieldness within the 44 disciplines (see the complete set of figures in our online Appendix B).\footnote{Online appendix is accessible under \url{https://gitlab.ethz.ch/raox/science-clf/-/blob/main/result_tables/online_appendix.pdf}.} In Figure \ref{fig:scatterplot_cs_econ}, we illustrate the said patterns for two disciplines that are sufficiently different, namely computer science and economics. 

In computer science, almost all fields have balanced demand and supply towards other fields within the discipline. Most of the fields in computer science have a high degree of interfield citations with one exception, applied computing (0-4).\footnote{We will see subsequently that applied computing  has a high interdisciplinarity score to fields outside of computer science (see Figures \ref{fig:f718_row} and \ref{fig:f718_col}), while having a low interfieldness score within its own discipline.}
In economics, we see a dispersion of demand and supply in fields, with fields like microeconomics (1-0), history of economic thought (1-7), economic systems (1-12) in the right upper corner having more references from outside fields and a high interfieldness score. 

%\textbf{XXX LET'S SUBSTITUTE THAT XXX. Let us examine the correlations between $\mathcal{I}_{dd}^$ and $\mathcal{O}_{dd}^$ without diagonal elements in all disciplines. 
% We use $\mathcal{I}_{dd}^*$ and $\mathcal{O}_{dd}^*$ to represent the matrices $\mathcal{I}_{dd}$ and $\mathcal{O}_{dd}$ without diagonal elements, respectively. We compute the Pearson correlation score ($r_d$) for $k^{out^*}_{Cdd}$ and $k^{in^*}_{Cdd}$ in each discipline shown in column (1) of Table \ref{tab:corr-interfield}. That is for each field within one discipline in $\mathcal{I}_{dd}$ and $\mathcal{O}_{dd}$, we compute \(\frac{\kappa^{inter}_{Cdd}} {\kappa^{total}_{Cdd}}\).  Focusing again on a comparison of computer science and economics, Here we observe, although the correlation scores are in similar scale, the two disciplines show different interfieldness as seen in Figure \ref{fig:scatterplot_cs_econ}.} 

%\textbf{XXX NEW XXX.} 
Let us use $\mathcal{I}_{dd0}-\mathcal{O}_{dd0}=-\mathcal{D}_{dd0}$ to denote the unnormalized net citation inflow matrix and $\mathcal{I}_{dd0}+\mathcal{O}_{dd0}$ to denote the unnormalized total citation flow matrix. Finally, let $\lvert -\mathcal{D}_{dd0} \rvert$ denote the absolute net in- or outflows of citations per field in discipline $d$. $(-\mathcal{D}_{dd0})\iota_{F_d}$ is a vector of (positive or negative) in-citation flows. Using $(\mathcal{I}_{dd0}\circ I_{F_d})$ as the $F_d\times F_d$ diagonal matrix of unnormalized intra-field citations in $d$, $k^{total}_{Cdd0}=(\mathcal{I}_{dd0}+\mathcal{O}_{dd0}-(\mathcal{I}_{dd0}\circ I_{F_d}))\iota_{F_d}$ is the vector of total in- and out-citations, avoiding intra-field citations to be counted twice. Then, the following scores can be defined for each field. The vector of intra-field citation scores within the discipline $d$ is $\varsigma^{intra}_d=\iota_{F_d}-diag(k^{total}_{Cdd0})^{-1}(\lvert -\mathcal{D}_{dd0} \rvert)\iota_{F_d}$. The elements of the latter are bounded between zero and one, and they are larger for fields with higher intra-field citation scores relative to all intra-disciplinary citations of all the fields demand and supply. $\varsigma^{unbal}_d=diag(k^{total}_{Cdd0})^{-1}(-\mathcal{D}_{dd0})\iota_{F_d}$ is a score that is bounded by $(-1,1)$ and indicates the relative degree of in- over out-citations per field. $\varsigma^{intra}_d$ and $\varsigma^{unbal}_d$ can be simply or field-citation-weighted averaged for each discipline $d$.

%% Newly added text, table tba
We report on these results in Table \ref{tab:sigma-unbal} for a simple average of $\varsigma^{intra}_d$ in column (3) and $\varsigma^{unbal}_d$ in column (4) across all fields within one discipline $d$.\footnote{The full result table of $\varsigma^{intra}_d$ (column (3)) and $\varsigma^{unbal}_d$ (column (4)) on each field is accessible under \url{https://gitlab.ethz.ch/raox/science-clf/-/blob/main/result_tables/interfieldness_tables_appendix/output_sigma.csv}.} A higher score of $\varsigma^{intra}_d$ indicates higher average intra- rather than inter-field contributions within a discipline. A more negative (positive) score of $\varsigma^{unbal}_d$ indicates that the average field in a discipline supplies (demands) more citations than it demands (supplies) from or to other fields within the respective discipline.\footnote{Note that $\varsigma^{unbal}_d$ does not mechanically sum to zero for a discipline because of the asymmetry of $\mathcal{I}$.}

The table indicates that disciplines ``Pure mathematics'', ``Logic'', ``Military science", and ``Journalism'' exhibit particularly high intra-field scores, whereas ``Economics'', ``Physics'', ``Family Studies'', and ``Systems science'' have above average intra-field citation scores. Furthermore, the results suggest that ``Physics'', ``Area studies'', ``Family studies'', and ``Systems science'' (most of those apearing with low intra-field scores above) have relatively large intradisciplinary net donors of citations (the smallest negative $\varsigma^{unbal}_d$ values in column (4) of Table \ref{tab:sigma-unbal}), whereas ``Computer science'', ``Architecture'', ``Performing arts'', and ``Transportation studies''have relatively large net recipients (the largest positive $\varsigma^{unbal}_d$ in column (4) of Table \ref{tab:sigma-unbal}).

\begin{table}[!th]
\centering
\caption{Interfieldness scores per discipline $d$ within and across disciplines. \\ \tiny [\textit{Within}-discipline] column (3): intra-field scores $\varsigma_{d}^{intra}$, column (4): relative in- over out citation scores $\varsigma_{d}^{unbal}$. [\textit{Across}-discipline] column (5): in- and out-citation scores overlap in interdisciplinary citation $\varsigma_{d}^{inter}$, \\column (6): relative in- over out-citation scores $\varsigma_{\neg d}^{unbal}$. \\ We mark the top 4 highest scores per column in \textbf{bold} and \underline{underline} the four lowest scores. }

\label{tab:sigma-unbal}
\resizebox{\linewidth}{!}{
\begin{tabular}{cccrrrr}
\toprule
\textbf{Discipline label} & \textbf{Discipline} & \textbf{$\varsigma^{intra}_d$} & \textbf{$\varsigma^{unbal}_d$} & \textbf{$\varsigma^{inter}_{d}$} & \textbf{$\varsigma^{unbal}_{\neg d}$} & \\
(1) & (2) & (3) & (4) & (5) & (6) \\
\midrule
0     & Computer science       & 0.8846          & \textbf{0.0675} & 0.8558          & \textbf{0.0880} \\
1     & Economics              & {\ul 0.8231}    & 0.0054          & {\ul 0.7938}    & {\ul -0.0948}   \\
2     & Pure mathematics       & \textbf{0.9933} & -0.0001         & \textbf{0.9575} & -0.0234         \\
3     & Physics                & {\ul 0.8357}    & {\ul -0.0581}   & 0.8216          & {\ul -0.0921}   \\
4     & Agriculture            & 0.9315          & -0.0299         & 0.8630          & 0.0208          \\
5     & Anthropology           & 0.9758          & -0.0180         & 0.9366          & 0.0089          \\
6     & Architecture           & 0.8390          & \textbf{0.0888} & {\ul 0.8004}    & \textbf{0.1167} \\
7     & Area studies           & 0.8729          & {\ul -0.0682}   & 0.8532          & -0.0104         \\
8     & Divinity       & 0.9130          & -0.0283         & 0.8491          & {\ul -0.0754}   \\
9     & Earth science          & 0.9409          & -0.0147         & 0.9114          & -0.0064         \\
10    & Environmental science  & 0.9582          & 0.0001          & 0.8855          & 0.0655          \\
11    & Ethnic studies         & 0.9443          & -0.0165         & 0.9083          & -0.0541         \\
12    & Family studies         & {\ul 0.7338}    & {\ul -0.0969}   & {\ul 0.5721}    & \textbf{0.2574} \\
13    & Gender studies         & 0.9378          & -0.0030         & 0.9021          & 0.0438          \\
14    & Geography              & 0.9159          & 0.0236          & 0.8745          & 0.0554          \\
15    & Human performances     & 0.8906          & 0.0270          & 0.8712          & 0.0424          \\
16    & Law                    & 0.9313          & -0.0302         & 0.9000          & -0.0534         \\
17    & Library science        & 0.9555          & 0.0283          & 0.9056          & 0.0653          \\
18    & Linguistics            & 0.9040          & -0.0141         & 0.8904          & -0.0404         \\
19    & Literature             & 0.9253          & -0.0215         & 0.8833          & -0.0516         \\
20    & Organizational studies & 0.9156          & -0.0117         & 0.9003          & 0.0132          \\
21    & Performing arts        & 0.9227          & \textbf{0.0296} & 0.8535          & \textbf{0.1051} \\
22    & Philosophy             & 0.9708          & 0.0010          & 0.9468          & -0.0117         \\
23    & Political science      & 0.9469          & 0.0002          & 0.9054          & 0.0465          \\
24    & Public administration  & 0.9346          & 0.0055          & 0.9085          & 0.0245          \\
25    & Religious studies      & 0.9507          & -0.0123         & 0.9158          & -0.0250         \\
26    & Social work            & 0.9617          & 0.0060          & 0.9332          & 0.0253          \\
27    & Space science          & 0.9701          & 0.0049          & 0.8942          & {\ul -0.0760}   \\
28    & Systems science        & {\ul 0.8328}    & {\ul -0.0487}   & {\ul 0.7869}    & 0.0284          \\
29    & Transportation studies & 0.9453          & \textbf{0.0296} & 0.9194          & 0.0416          \\
30    & Visual arts            & 0.9770          & -0.0097         & 0.9064          & -0.0238         \\
31    & Logic                  & \textbf{0.9851} & 0.0024          & \textbf{0.9623} & -0.0060         \\
32    & Education              & 0.9325          & 0.0158          & 0.9040          & 0.0538          \\
33    & Business               & 0.9362          & 0.0067          & 0.8959          & 0.0681          \\
34    & Chemistry              & 0.9083          & -0.0033         & 0.9112          & -0.0332         \\
35    & Engineering            & 0.9301          & -0.0267         & 0.8929          & 0.0597          \\
36    & History                & 0.9473          & -0.0183         & 0.9105          & -0.0130         \\
37    & Medicine               & 0.9305          & 0.0205          & 0.9050          & 0.0271          \\
38    & Military science       & \textbf{0.9820} & -0.0068         & \textbf{0.9789} & -0.0152         \\
39    & Psychology             & 0.9460          & -0.0059         & 0.9445          & -0.0073         \\
40    & Sociology              & 0.9106          & -0.0239         & 0.9064          & -0.0378         \\
41    & Journalism             & \textbf{0.9919} & 0.0032          & 0.9513          & 0.0179          \\
42    & Applied mathematics    & 0.9641          & -0.0136         & \textbf{0.9578} & -0.0303         \\
43    & Biology                & 0.8914          & 0.0245          & 0.9149          & -0.0380                \\
\bottomrule
\end{tabular}}

\end{table}

\subsubsection{Across-Discipline Interfieldness (Interdisciplinarity)} \label{sec:across-discipline}
We next consider the interfieldness of the 718 fields \textit{across} disciplines. Therefore, the focus in this section is entirely on interdisciplinarity. Before scrutinizing on individual fields, let us consider the scores at the level of aggregate disciplines. 

It will be useful to define the block-diagonal matrix of unnormalized in-citations $\mathcal{B}_0=diag^D_{d=1}(\mathcal{I}_{dd0})$ as well as the matrices $\mathcal{I}^*_0=\mathcal{I}_0-\mathcal{B}_0$ and $\mathcal{O}^*_0=\mathcal{I}^{*\prime}_0$. The latter two matrices exclude the intra-disciplinary citation blocks $\mathcal{I}_{dd0}$ visualized in Figure \ref{fig:interfieldness-vis}. Upon using $\mathcal{I}$ and $\mathcal{O}$ and $\mathcal{B}=diag^D_{d=1}(\mathcal{I}_{dd})$, we can obtain the normalized $\mathcal{I}^*$ and $\mathcal{O}^*$ analogously to $\mathcal{O}^*_0=\mathcal{I}^{*\prime}_0$. Note that their rows do not sum up to unity, but to less than unity, depending on the relative weight of intradisciplinary citations.

\begin{figure}[!t]
    \centering
    \includegraphics[width=1\linewidth]{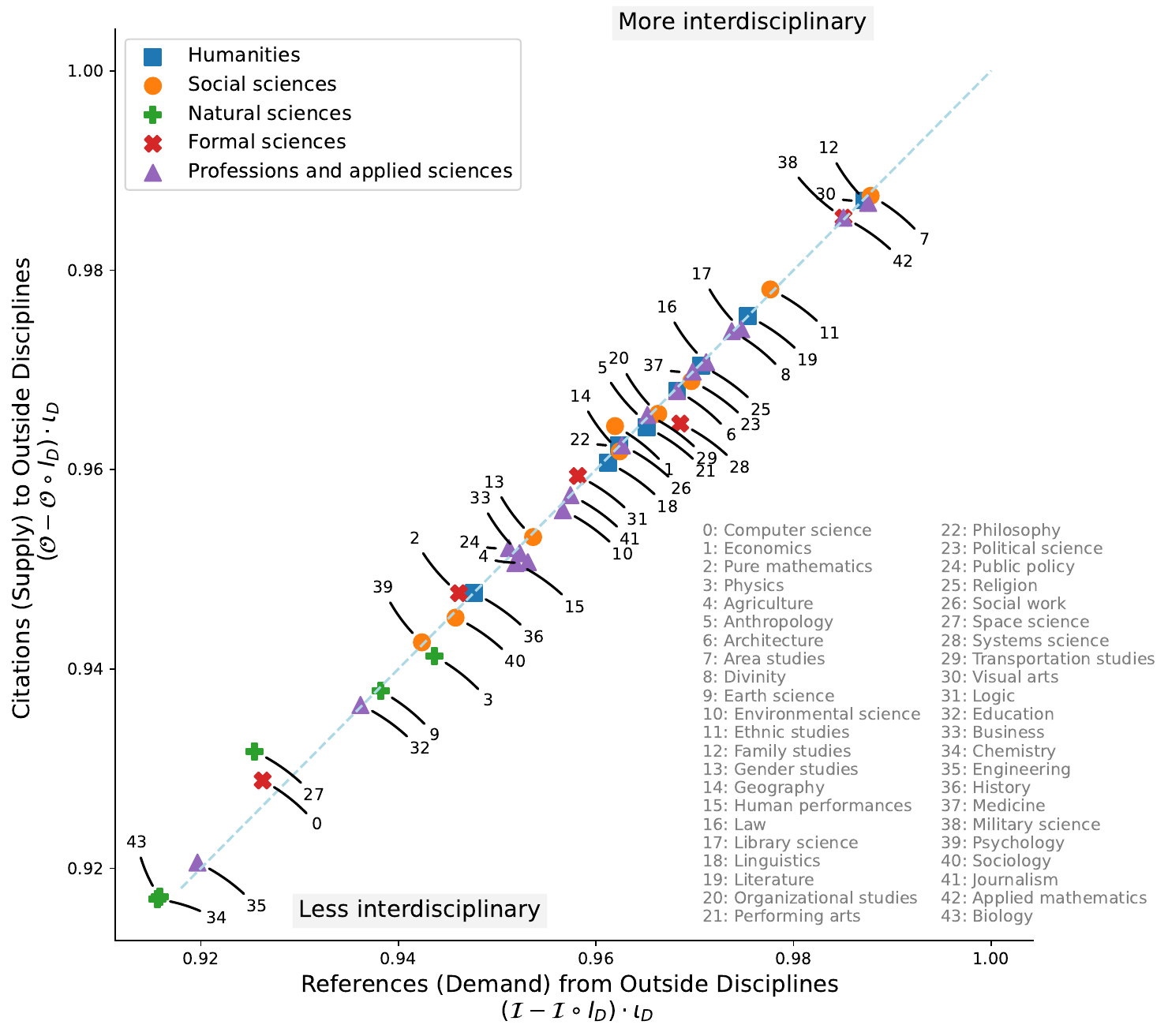}
    \caption{Degrees of interdisciplinarity by demand from and supply to outside disciplines in 44 disciplines.}
    \label{fig:scatterplot}
\end{figure}

Similar to the exercise in Figure \ref{fig:scatterplot_cs_econ}, we plot the demand from and supply to outside disciplines in terms of citations in Figure \ref{fig:scatterplot}. The underlying metrics are based on the $D$ row sums of $44\times 44$ (aggregated) discipline-level counterpart matrices to $\mathcal{I}^*$ and $\mathcal{O}^*$. The aggregated matrices obtain one pair of (in- and out-citation interdiscipline) values for each of the 44 disciplines. Similar to Figure 3 in \cite{van2015interdisciplinary}, we show the degrees of interdisciplinarity at the level of discipline in Figure \ref{fig:scatterplot}. 

In the figure, more interdisciplinary disciplines are found in the top-right area and the less interdisciplinary ones are found in the bottom-left area of the figure. The position of a discipline is determined by two factors: the extent to which its research output is cited by disciplines outside its own (on the abscissa), and the extent to which those outside disciplines are cited by that discipline (on the ordinate). As with the fields-based matrix, the scores are lower than unity, because the diagonal (here a scalar, at the level of fields a matrix) is subtracted. Figure \ref{fig:out_normalized} indicates that the disciplines vary significantly in terms of their interdisciplinarity scores. At the low interdisciplinarity end we find ``Biology'' and ``Chemistry'', while at the high end we find ``Area studies'' and ``Family studies''. 

For better illustration, we group the disciplines by the Wikipedia discipline classification depicted in Figure \ref{fig:wiki}. There are five broad categories: ``Humanities", ``Social sciences", ``Natural sciences", ``Formal sciences", and ``Professions and applied sciences". The visualization suggests that there is a lower degree of interdisciplinarity in the ``Natural sciences" (\textcolor{green}{\textbf{+}}). ``Professions and applied sciences" (\textcolor{violet}{\large $\rotatebox[origin=c]{90}{$\blacktriangleright$}$}) have a large variation in terms of interdisciplinarity. In that category, ``Engineering and technology" (35) shows the least demand and supply from and to outside disciplines, while ``Education" (32) has a high interdisciplinary demand and supply. Apart from ``Computer science" (0), the other ``Formal sciences" (\textcolor{red}{\textbf{$\times$}}) have a high interdisciplinarity. Disciplines in the ``Humanities" (\textcolor{blue}{\rule{1em}{1em}}) have a relatively high demand and supply from and to outside disciplines, and so do disciplines in ``Social sciences" (\textcolor{red}{\textbf{\large $\bullet$}}).  

These results only partially corroborate the ones in Figure 3 in \cite{van2015interdisciplinary}. There as well as here the ``Social sciences" display a high degree of interdisciplinarity and ``Physics" (3) has a low one. Some fields such as ``Anthropology" (4) or ``Psychology" (39) have similar positions w.r.t.~other disciplines between \cite{van2015interdisciplinary} and our study. Yet, some other fields (``Biology" (43) and ``Applied mathematics" (42)) show an opposite positioning in comparison to \cite{van2015interdisciplinary}.\footnote{``Biology" in Figure 3 of \cite{van2015interdisciplinary} has a wide variation in interdisciplinarity, with ``General biology" on the upper right corner and the remaining fields scattered in the middle spectrum of distribution.} Since the sample in \cite{van2015interdisciplinary} is based on only 35 million articles in 2001-2010 in the Web of Science of 14 major conventional disciplines and 143 fields, we cannot draw decisive conclusions based on the differences. However, any differences may suggest there being interesting results to emerge from tracking longitudinal changes of interdisciplinarity in larger disciplinary cross sections.

\begin{figure}[!ht]
    \centering
    \includegraphics[width=1\linewidth]{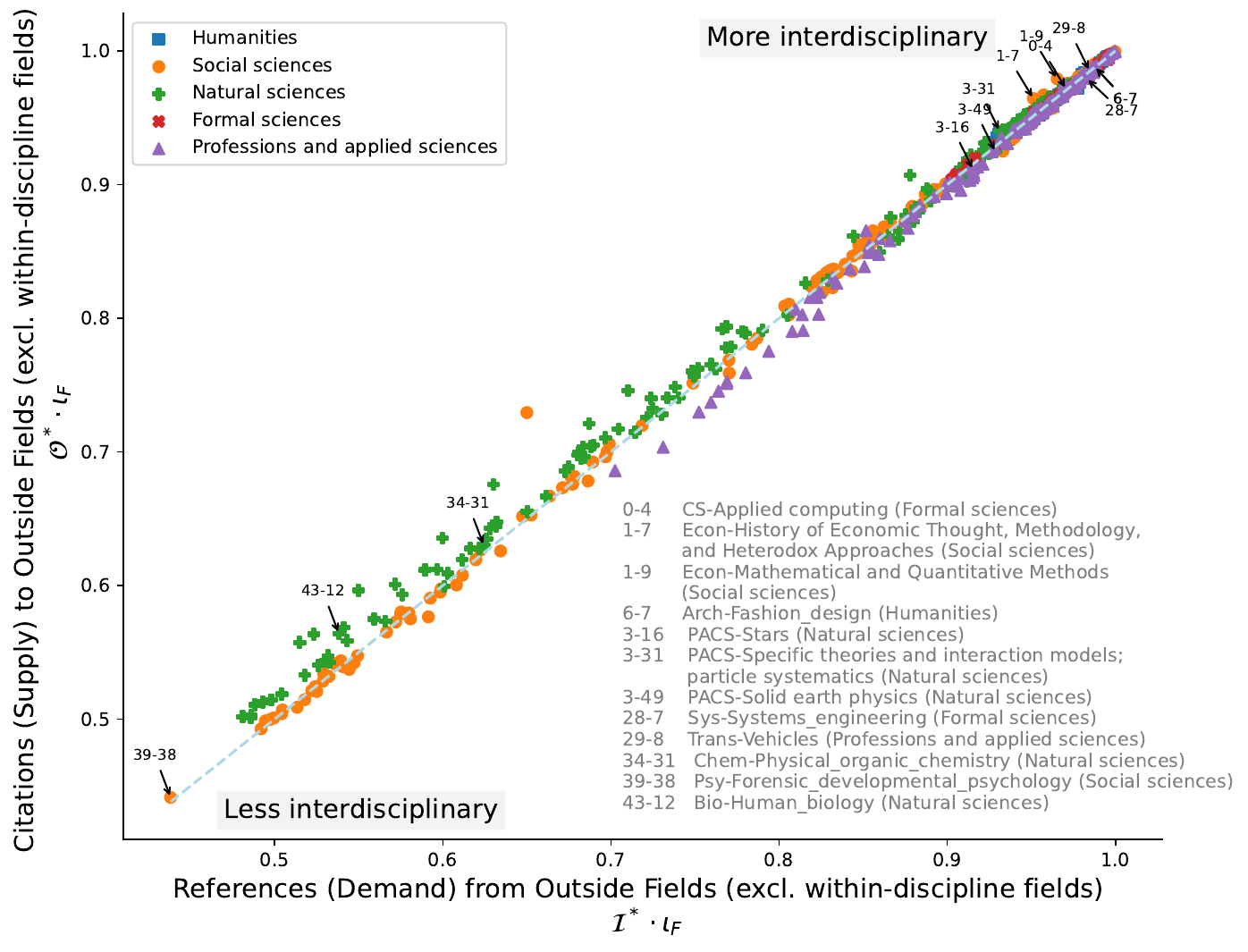}
    \caption{Degrees of interdisciplinarity by demand from and supply to outside fields in 718 fields, excl.~within-discipline fields.}
    \label{fig:scatterplot_718}
\end{figure}

Next we focus on individual fields in the 44 disciplines, of which there are $F=718$. We perform the same exercise as above for the 718 individual fields to measure their interdisciplinarity scores in Figure \ref{fig:scatterplot_718}, \textit{excluding} fields within each discipline.  The underlying metrics are now based on the $F$ row sums of $\mathcal{I}^*$ and $\mathcal{O}^*$, obtaining one pair of values for each of the 718 fields. Again, we augment the analysis with results grouped by the five broad categories of disciplines used above. The respective analysis suggests distinctive patterns for the fields in different categories. (1) Ones in the ``Natural sciences'' (\textcolor{green}{\textbf{+}}) and ``Social sciences" (\textcolor{red}{\textbf{\large $\bullet$}}) have the highest variability of interdisciplinarity. (2) Fields in ``Professions and applied sciences" (\textcolor{violet}{\large $\rotatebox[origin=c]{90}{$\blacktriangleright$}$}), ``Humanities" (\textcolor{blue}{\rule{1em}{1em}}) , and ``Formal sciences" (\textcolor{red}{\textbf{$\times$}}) have a relatively high average degree interdisciplinarity. 

We mark fields that are in various spectrums of the distribution for a better comparison between the two visualizations. Clearly, we see some bimodality for the fields we single out: some are highly dependent on fields within their home disciplines (39-38, 43-12, 34-31), while others are very interdisciplinary (6-7, 28-7). 

Let us consider $\varsigma^{inter}_{d}$ and $\varsigma^{unbal}_{\neg d}$, with $\neg d$ denoting the fields that are outside of a discipline $d$. These two are $F_d\times 1$ vectors for discipline $d$, which are defined as the discipline-$d$ specific $F_d\times 1$ subvectors of:
\begin{eqnarray}
\varsigma^{inter}_d&=&\iota_{F}-diag(k^{total}_{F})^{-1}(\lvert \mathcal{I}^*_0-\mathcal{O}^*_0 \rvert)\iota_{F}, \\
\varsigma^{unbal}_{\neg d}&=&diag(k^{total}_{F})^{-1}(\mathcal{I}^*_0-\mathcal{O}^*_0)\iota_{F},
\end{eqnarray}
where $k^{total}_{F} = (\mathcal{I}_0+\mathcal{O}_0-(\mathcal{O}_0 \circ I_F)) \iota_F$. 
The term $\varsigma^{inter}_{d}$ captures the in- and out-citation overlap in interdisciplinary citation scores per field in discipline $d$ relative to all others. $\varsigma^{unbal}_{\neg d}$ measures the relative degree of in- over out-citations for the average pair of fields in one discipline relative to all others.

The results are reported in columns (5) and (6) of Table \ref{tab:sigma-unbal} by taking the simple average of all fields within discipline $d$.\footnote{The full result table of $\varsigma^{inter}_{d}$ (column (5)) and $\varsigma^{unbal}_{\neg d}$ (column (6)) on each field is accessible at \url{https://gitlab.ethz.ch/raox/science-clf/-/blob/main/result_tables/interfieldness_tables_appendix/output_sigma_wo_dfields.csv}.}  In the interdisciplinary setting, we find that ``Pure mathematics", ``Logic", ``Military science", and ``Applied mathematics" have particularly high in-out-citation overlap scores $\varsigma^{inter}_d$, and ``Architecture" (6), ``Family studies" (12), ``Systems Science" (28) have above average in- and out-citation overlapping scores. These disciplines have both a relatively high $\varsigma^{intra}_d$ and $\varsigma_d^{inter}$ compared to other disciplines. These findings are consistent with the discipline-level analysis in Figure \ref{fig:scatterplot}. They are situated in the middle-upper spectrum of the distribution, which indicates a relative high demand from and supply to the outside disciplines.

The ranking of $\varsigma^{unbal}_{\neg d}$ of some disciplines has changed a lot compared to $\varsigma^{unbal}_d$ within the discipline. In interdisciplinary settings, ``Family studies" has become a large net recipient (the largest $\varsigma^{unbal}_{\neg d}$), while ``Economics" has become a net donor (negative $\varsigma^{unbal}_{\neg d}$). 

Overall, by considering fields as the unit of analysis in studying interdisciplinarity, it is possible to attribute metrics to fields and consider the distribution within and across disciplines in finer granularity. This approach provides a more comprehensive understanding of research links across disciplines, eventually even over time and in geographical space.

\section{Conclusion and Future Work}
In this paper, we have devised a three-level hierarchical classification system for scientific publications based on state-of-the-art deep learning methods. This system supports multi-label classifications in both single-label and multi-label settings. We have enabled a modularized classification system that copes with a large and increasing number of publications and supports quick update of submodels. We have conducted numerous experiments to test the efficiency of our system. Moreover, we have developed analytics and metrics in measuring interfieldness within and across disciplines on the level of field. We provide this platform and data to the research community and invite joint efforts to enable an efficient ecosystem in the classification of scholarly publications and the analysis of interdisciplinarity. 

\begin{figure}[!h]
    \centering
    \includegraphics[width=\textwidth]{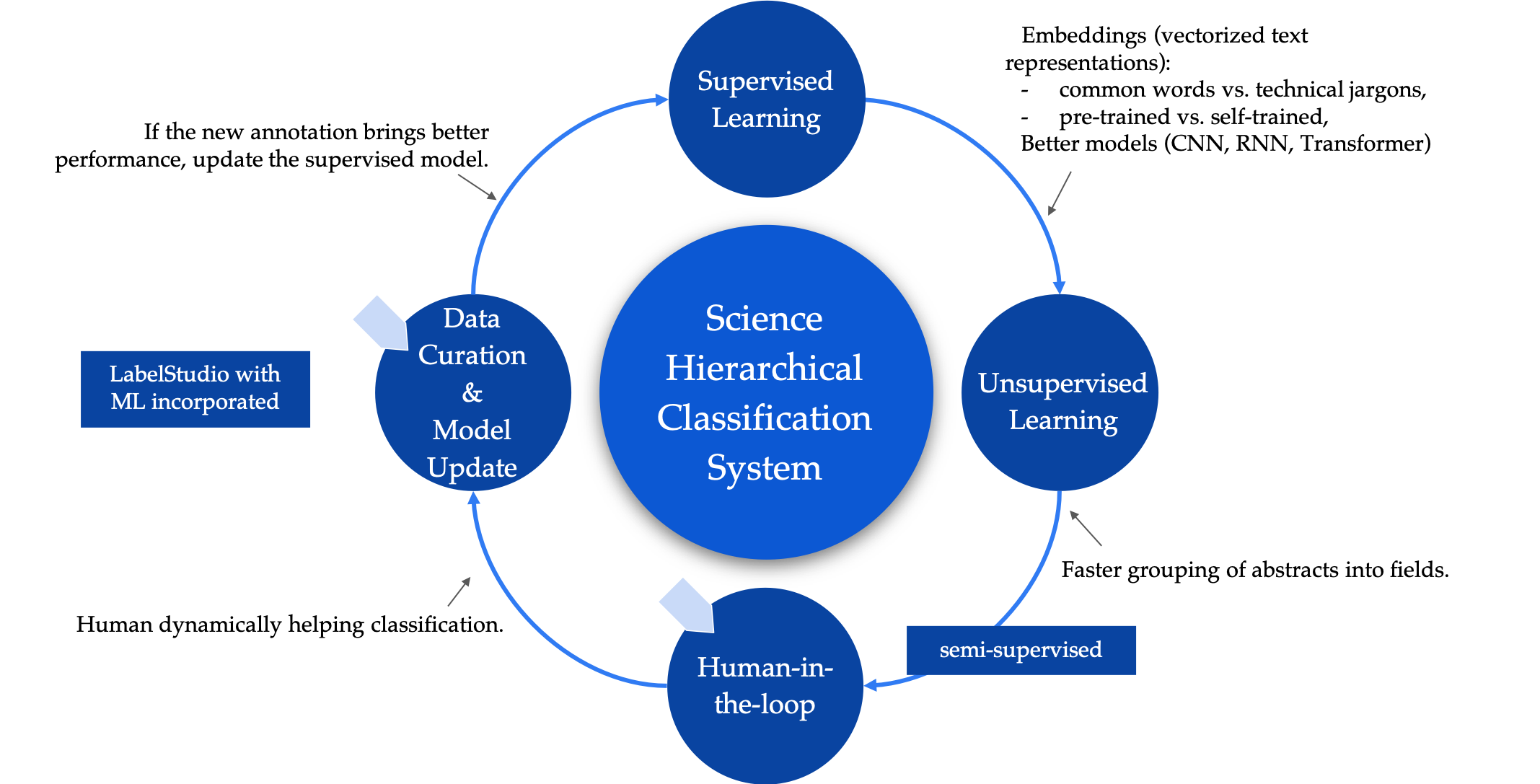}
    \caption{Science hierarchical classification lifecycle.}
    \label{fig:cycle}
\end{figure}

As introduced in Section \ref{sec:intro}, this project is part of a larger research agenda which aims to tackle multifaceted problems in archiving and organizing exponentially growing scientific publications. As a next step shown in Figure~\ref{fig:cycle}, we would like to involve \textit{human-in-the-loop} (like human annotations for model improvement) in crowd-sourcing label quality improvement and, in exchange, provide an even higher quality database of scholarly publications. Especially, now that Microsoft Academic Services is no longer available, we are looking to provide a brand new competitive solution to existing players in the market, deploying machine learning models into an interactive web front-end using Label Studio \citep{LabelStudio}, an open-source data labeling tool. To this end, we have already developed SAINE \citep{rao2023saine}, a Scientific Annotation and Inference Engine of Scientific Research, to better understand classification results.

In future work, as an extension of the interfieldness analysis, we plan to use the entire graph (after inference using the models) using OpenAlex to compute the statistics reported in Section \ref{sec:interdisciplinarity}. In addition, we are also working on reproducing the \textit{ogbn}-MAG benchmark \citep{hu2020open} using our own field classifications. 
\clearpage
%citation recommender system
%Is there a way we can better utilize the data and search mechanism at work behind those products? Could we generate a hierarchical structure with the search results where the article contents are associated with some discipline/field/category?

\backmatter

\bmhead{Acknowledgments}

We thank Ms.~Piriyakorn Piriyatamwong for her technical support to our project. The authors also thank the Microsoft Academic Service and especially Charles Huang for providing instructions to download the data. 
%upon acceptance, colleagues at CAE, DS3LAB, Systems Group

\section*{Statements and Declarations}
\begin{itemize}
\item \textbf{Funding}

Peter Egger and the Chair of Applied Economics acknowledge the support of the Department of Management, Technology, and Economics at ETH Zurich. % Peter
Ce Zhang and the DS3Lab gratefully acknowledge the support from the Swiss State Secretariat for Education, Research and Innovation (SERI) under contract number MB22.00036 (for European Research Council (ERC) Starting Grant TRIDENT 101042665), the Swiss National Science Foundation (Project Number 200021\_184628, and 197485), Innosuisse/SNF BRIDGE Discovery (Project Number 40B2-0\_187132), European Union Horizon 2020 Research and Innovation Programme (DAPHNE, 957407), Botnar Research Centre for Child Health, Swiss Data Science Center, Alibaba, Cisco, eBay, Google Focused Research Awards, Kuaishou Inc., Oracle Labs, Zurich Insurance, and the Department of Computer Science at ETH Zurich. % Ce
\item \textbf{Conflict of interest/Competing interests} 

ETH Zurich Domain; University of Zurich, Department of Computational Linguistics: \url{https://www.cl.uzh.ch/en.html}. 

\item \textbf{Ethics approval }

Not applicable.

\item \textbf{Availability of data and materials}

We plan to make the data publicly available (raw data and intermediary data) upon acceptance of the manuscript. Due to the sheer size of raw and intermediary data, we will upload the LMDB instances for each discipline to Zenodo or Mendeley Data. 

\item \textbf{Code availability }

The code base has been made available under \url{https://gitlab.ethz.ch/raox/science-clf}.

\item \textbf{Authors' contributions (14 roles in CRediT)}

(1) Conceptualization: Peter Egger, Susie Xi Rao, Ce Zhang. 
(2) Data curation: Susie Xi Rao. 
(3) Formal analysis: Peter Egger, Susie Xi Rao.
(4) Funding acquisition: Peter Egger, Ce Zhang, Susie Xi Rao.
(5) Investigation: Susie Xi Rao, Peter Egger.
(6) Methodology: Susie Xi Rao, Ce Zhang, Peter Egger. 
(7) Project administration: Peter Egger, Ce Zhang, Susie Xi Rao.
(8) Resources: Peter Egger, Ce Zhang, Susie Xi Rao.
(9) Software: Susie Xi Rao.
(10) Supervision: Peter Egger, Ce Zhang.
(11) Validation: Susie Xi Rao, Peter Egger.
(12) Visualization: Susie Xi Rao, Peter Egger.
(13) Writing - original draft: Susie Xi Rao, Peter Egger.
(14) Writing - review \& editing: Peter Egger, Susie Xi Rao, Ce Zhang. 
\end{itemize}

\clearpage

\begin{appendices}\label{sec:appendix}

\section{Case Study of Economics, Computer Science and Mathematics}\label{app:case-study-econ-cs}
To obtain the most accurate ground truth labels for every hierarchical level, we decided to proceed with papers that are archived according to systematic hierarchical classification code systems. As such, we restrict our attention to papers from the following three sources as the first case study. 

\begin{enumerate}
    \item Association for Computing Machinery (ACM), a peer-reviewed journal for computer science.
    \item Journal of Economic Literature (JEL), a peer-reviewed journal for economics.
    \item Mathematics Subject Classification (MSC), a classification system for mathematical publications.
\end{enumerate}

By restricting ourselves to three disciplines, we show proof of concepts of how the hierarchical classification performs. We list the class distribution in the three datasets in Table~\ref{table:mag-case}.

\begin{table}[h!]
\resizebox{\columnwidth}{!}{
\begin{tabular}{l|r|r|r}
\toprule
\textbf{Discipline (L1)}        & \textbf{Number of documents} & \textbf{\begin{tabular}[c]{@{}c@{}}Number of \\ disciplines (L2)\end{tabular}} & \textbf{\begin{tabular}[c]{@{}c@{}}Number of \\ disciplines (L3)\end{tabular}}     \\ \midrule
ACM (Computer Science)           & 11,637,219       & 13                  & 13                                        \\ \midrule
JEL (Economics)      & 8,439,655                        & 15    & 47                                   \\ \midrule
MSC (Mathematics)       & 10,214,588                        & 55     & 252                                  \\ \midrule
\textbf{Total}         & \textbf{30,291,462}              & \textbf{83} & \textbf{312}                             \\ \bottomrule
\end{tabular}
}
\caption{Summary of MAG datasets used in our case study.}
\label{table:mag-case}
\end{table}

\subsection{Topic modelling}\label{app:topic-model}
We have tested topic modeling in both supervised \citep{mcauliffe2007supervised} and unsupervised fashions \citep{blei2003latent} by first discovering topics of 55 disciplines using their textual descriptions on the Wikipedia page ``List of academic fields". Unfortunately, the topic words generated via topic modelling failed to capture the nuance between fields (class confusability). Using these lists of words to represent a field is too coarse to link a publication in MAG to a field. We list four subfields (``Biophysics", ``Molecular biology", ``Structural Biology", ``Biochemistry") below that share a similar set of topic words / topic nouns using a supervised topic model. The complete list of topic words for all fields and their field description could be found \href{here}{https://gitlab.ethz.ch/raox/science-clf/-/tree/main/Wikipedia_hierarchy} in our project repository.

\lstset{language=XML}
\begin{lstlisting}
<topic>Biophysics</topic>
<TopicWords>biophysics, biology, -, molecular, biological, 
biophysical, study, department, structure, technique,
system, research, quantum, physic, physiology, chemistry,
protein, biochemistry, interaction, field, medicine, physical,
include, feynman, biomolecular, model, science, apply,
biophysicist, journal, mathematics, experimental, effort,
complex, population, cell, structural, see, molecule, use,
example, microscopy, application, neural, dynamics, computer,
cellular, member, medical, neutron</TopicWords>
<TopicNouns>biophysics, biology, department, study, structure,
technique, system, research, physiology, physic, quantum,
field, biochemistry, medicine, chemistry, interaction,
protein, model, science, journal, biophysicist, feynman,
application, effort, population, mathematics, molecule,
example, cell, computer, dynamics, microscopy, brain, variety,
membrane, nanomedicine, event, alignment, list, complex, gene,
tissue, spectroscopy, idea, machine, society, kinetics,
neutron, physicist, network</TopicNouns>


<topic>Molecular_biology</topic>
<TopicWords>isbn, edition, garland, biology, molecular, base,
biochemistry, rd, pound, nd, link, curlie, external, dmoz,
dna, technique, blot, protein, cell, rna, study, gel,
molecule, gene, probe, pcr, one, size, specific, use,
electrophoresis, expression, array, label, sample, membrane,
interest, sequence, spot, clon, separate, allow, southern,
different, field, via, function, fragment, transfection,
enzyme</TopicWords>
<TopicNouns>isbn, edition, garland, biology, biochemistry, rd,
pound, nd, link, curlie, dmoz, dna, technique, protein, cell,
rna, gel, molecule, pcr, study, gene, size, expression,
electrophoresis, interest, sequence, sample, membrane, spot,
array, blot, probe, field, science, target, function, tissue,
enzyme, transfection, fragment, genetics, interaction,
process, organism, site, restriction, base, time, reaction,
basis</TopicNouns>

<topic>Structural_Biology</topic>
<TopicWords>structure, structural, protein, biology,
molecular, molecule, see, method, model, macromolecule,
function, biologist, shape, acid, make, biochemistry,
biological, cell, tertiary, primary, light, use, native,
small, membrane, study, state, prediction, scattering,
resonance, researcher, electron, spectroscopy, aspect,
physical, deduce, hydrophobicity, integral, amino, predict,
alteration, accurate, become, diverse, complement, highly,
understanding, base, topology, approach</TopicWords>
<TopicNouns>structure, protein, biology, molecule, model,
method, function, macromolecule, biologist, light, cell,
biochemistry, acid, electron, researcher, use, spectroscopy,
resonance, prediction, scattering, shape, study, state,
membrane, library, link, magazine, nature, subunit, europe,
cooperativity, pattern, example, datum, reference,
understanding, year, silico, amino, topology, biophysics,
hydrophobicity, aspect, analysis, journal, sequence, density,
chaperonin, bank, bioinformatics</TopicNouns>

<topic>Biochemistry</topic>
<TopicWords>acid, molecule, amino, biochemistry, protein,
form, call, one, two, cell, carbon, glucose, structure, group,
energy, study, enzyme, oxygen, process, biology, molecular,
reaction, biological, life, use, chain, example, glycolysis,
nucleic, organism, carbohydrate, make, living, convert, base,
atom, genetic, reduce, animal, monosaccharide, important, atp,
join, bond, chemistry, lipid, human, function, another,
chemical</TopicWords>
<TopicNouns>acid, molecule, amino, biochemistry, protein,
form, cell, carbon, glucose, structure, group, energy, study,
oxygen, enzyme, process, biology, reaction, life, chain,
example, glycolysis, organism, carbohydrate, living, atom,
animal, monosaccharide, atp, bond, chemistry, lipid, function,
chemical, monomer, sugar, base, pathway, component, role,
information, plant, water, nadh, adenine, ring, rna, residue,
cycle, gene</TopicNouns>
        
\end{lstlisting}

Our trials of topic modelling have led us to supervised methods, i.e., classification. And our data are organized in a hierarchical fashion, i.e., one publication (its representation being its abstract) and its labels (\textit{discipline-field-subfield}). The objective of developing a hierarchical classification system is to leverage the hierarchical organization to create models and classify unlabeled test instances into one or more categories within the hierarchy.

\subsection{Hierarchical SVM}\label{app:h-svm}

Classification methods proven effective in hierarchical settings are multi-class support vector machine (SVM) \citep{sun2001hierarchical} and stacking SVM, i.e., ensemble of individual SVM classifiers \citep{kenji2017stacking}. We carried out experiments using hierarchical multi-class SVM on a benchmarking dataset WOS-46985 published in \cite{kowsari2017hdltex}. The classification is a two-level system with 13 first-level (L1) labels and 76 second-level (L2) labels and has achieved only 37\% of macro-F1 across all the L2 classes. 

%%% Comment out to speed up compilation for R&R

\subsection{Two-Level Classification}\label{app:2lsys}
As we see from the poor performance of traditional machine learning techniques in the trials, we need a better solution. \cite{kowsari2017hdltex} has suggested a success of a hierarchical two-level classification system. The idea has been attractive not because the hype of the belief that every application should have a deep learning component which just magically makes the performance better, but because we can leverage the hierarchy structure in the data by linking the submodels on each level by the stacking of layers, and we can capture the probability distributions of multiple classes via the \textit{softmax} at the last layer of each classification component. We denote $p$ a publication, $D$ one discipline, $F_i$ one field in $D$. We obtain the unconditional probability $P(p \in D)$ from the first component in the system that classifies the disciplines and the conditional probability $ P(p \in F_i \mid p \in D) $ from the second component that classifies the fields. In \citet{piriMT}, a master thesis supervised by Susie Xi Rao, we have shown reasonable performances for a three-discipline classification (ACM, JEL and MSC) in both scalability and prediction accuracy.

\section{Training Setup of Our Classifier System}
\label{app:training-setup}
{We thank Ms.~Piriyakorn Piriyatamwong for her technical support to our project. Note that an ablation study of hyperparameters was presented in her master thesis \citep{piriMT} supervised by Susie Xi Rao. Hence, we took over the hyperparameters configured in this study and reported them in Appendix~\ref{app:training-setup}. }

Text Vectorization has the following set up:

\begin{itemize}
    \item Maximum length of text vectors: 200, which makes sense since abstracts are usually capped to 150-250 words.
    \item Split: by white space.
    \item Normalization: lower-casing and punctuation removals.
\end{itemize}

\noindent Keras models have the following set up:

\begin{itemize}
    \item Batch size: 1,024 unless the model dataset is too small, then the largest power of 2 smaller than the model dataset.
    \item Number of epochs: 1, as the performance is sufficiently good and not overfitted.
    \item Optimizer: Root Mean Square Propagation (RMSProp)\footnote{\href{http://www.cs.toronto.edu/~hinton/coursera/lecture6/lec6.pdf}{http://www.cs.toronto.edu/~hinton/coursera/lecture6/lec6.pdf}  (last accessed: May 23, 2022).} with learning rate 0.001. RMSProp accelerates the gradient descent process of optimizing the loss function.
    \item Train-test split ratio: 60\% training, 40\% test.
\end{itemize}

\noindent BERT uncased model has the following set up, following the recommendations from the original paper \citep{devlin-etal-2019-bert}.

\begin{itemize}
    \item Batch size: 16 or 32.
    \item Number of epochs: 2, 3, or 4.
    \item Optimizer: Adam optimizer with learning rate 2e-5, another computationally efficient, scalable optimizer with low memory requirement \citep{DBLP:journals/corr/KingmaB14}. 
    \item Train-test split ratio: 60\% training, 40\% test.
\end{itemize}

\section{Discipline-to-Coding and Field-to-Coding Mappings}
\label{app:discipline2coding-mapping}
We list here the discipline to coding mapping for readers' reference. The 718 field labels are accessible in our project repository under the link \url{https://gitlab.ethz.ch/raox/science-clf/-/blob/main/result_tables/interfieldness_tables_appendix/718_field_labels_master_copy.csv}.
\begin{multicols}{2}
\lstset{language=json}
\begin{lstlisting}
    "infk": 0,
    "econ": 1,
    "purem": 2,
    "phys": 3,
    "agri": 4,
    "anthro": 5,
    "arch": 6,
    "area": 7,
    "div": 8, 
    "earth": 9,
    "env": 10,
    "ethnic": 11,
    "fam": 12,
    "gender": 13,
    "geo": 14,
    "human": 15,
    "law": 16,
    "lib": 17,
    "ling": 18,
    "lit": 19,
    "org": 20,
    "perf": 21,
    "phil": 22,
    "pol": 23,
    "pub": 24,
    "rel": 25,
    "socw": 26,
    "space": 27,
    "sys": 28,
    "trans": 29,
    "vis": 30,
    "logic": 31,
    "edu": 32,
    "bus": 33,
    "chem": 34, 
    "eng": 35,
    "hist": 36,
    "med": 37,
    "mil": 38,
    "psy": 39,
    "soc": 40,
    "jour": 41,
    "appliedm": 42, 
    "bio": 43
\end{lstlisting}
\end{multicols}

%%%%%%%%%%%%%%%%commented out appendix




\end{appendices}


\bibliography{sn-bibliography}% common bib file


\end{document}